\documentclass[fleqn,usenatbib]{mnras}

\usepackage{newtxtext,newtxmath}
\usepackage[T1]{fontenc}
\usepackage[shortlabels]{enumitem}

\DeclareRobustCommand{\VAN}[3]{#2}
\let\VANthebibliography\thebibliography
\def\thebibliography{\DeclareRobustCommand{\VAN}[3]{##3}\VANthebibliography}

\usepackage{graphicx}	% Including figure files
\usepackage{amsmath}	% Advanced maths commands
\usepackage{calc}

\newcommand{\Msun}{\mathrm{M}_{\odot}}

\newcommand{\logMstellar}{\log(M_{\star}/\Msun)}
\newcommand*\diff{\mathop{}\!\mathrm{d}}

\makeatletter
\DeclareRobustCommand{\HI}{%
  \ensuremath{\text{H\check@mathfonts\fontsize\f@size\z@\scalebox{0.8}{\selectfont I}}}%
}
\makeatother

\title[How many stars form in galaxy mergers?]{How many stars form in galaxy mergers?}

\author[A. M. M. Reeves et al.]{
Andrew M. M. Reeves$^{1,2}$\thanks{E-mail: andrew.reeves@uwaterloo.ca},
Michael J. Hudson$^{1,2,3}$
\\
$^{1}$Department of Physics and Astronomy, University of Waterloo, Waterloo, ON N2L 3G1, Canada \\
$^{2}$Waterloo Centre for Astrophysics, University of Waterloo, Waterloo, ON N2L3G1, Canada \\
$^{3}$Perimeter Institute for Theoretical Physics, Waterloo, ON N2L 2Y5, Canada \\
}

\date{Accepted XXX. Received YYY; in original form ZZZ}

\pubyear{2023}

\begin{document}
\label{firstpage}
\pagerange{\pageref{firstpage}--\pageref{lastpage}}
\maketitle

\begin{abstract}
We forward model the difference in stellar age between post-coalescence mergers and a control sample with the same stellar mass, environmental density, and redshift. In particular, we use a pure sample of 445 post-coalescence mergers from the recent visually-confirmed post-coalescence merger sample identified by Bickley et al.\ and find that post-coalescence mergers are on average younger than control galaxies for $10<\logMstellar<11$. The difference in age from matched controls is up to 1.5~Gyr, highest for lower stellar mass galaxies. We forward model this difference using parametric star formation histories, accounting for the pre-coalescence inspiral phase of enhanced star formation using close pair data, and a final additive burst of star formation at coalescence. We find a best-fitting stellar mass burst fraction of $f_\mathrm{burst}=\Delta M_\star/M_{\star,\mathrm{merger}}=0.18 \pm 0.02$ for $10<\logMstellar<11$ galaxies, with no evidence of a trend in stellar mass. The modeled burst fraction is robust to choice of parametric star formation history, as well as differences in burst duration. The result appears consistent with some prior observationally-derived values, but is significantly higher than that found in hydrodynamical simulations. Using published Luminous InfraRed Galaxy (LIRG) star formation rates, we find a burst duration increasing with stellar mass, from 120--250~Myr. A comparison to published cold gas measurements indicates there is enough molecular gas available in very close pairs to fuel the burst. Additionally, given our stellar mass burst estimate, the predicted cold gas fraction remaining after the burst is consistent with observed post-coalescence mergers.
\end{abstract}


%%%%%%%%%%%%%%%%%%%%%%%%%%%%%%%%%%%%%%%%%%%%%%%%%%

%%%%%%%%%%%%%%%%% BODY OF PAPER %%%%%%%%%%%%%%%%%%

%%% %%% %%% %%% %%% %%% %%% %%% %%%
\section{Introduction} \label{sec:introduction}

Understanding the impact of the merging of galaxies is essential to formulating a complete picture of hierarchical galaxy evolution in the $\Lambda$CDM framework \citep{Kauffmann1993,Navarro1996,Somerville2015}. Mergers, both major and minor, are an intrinsic part of the build-up of stellar and dark matter mass to form the galaxies we see in today's universe, especially given their highly pronounced role early in the Universe's history \citep[e.g.][]{Conselice2003,Hopkins2010}. Mergers are not only additive, but also transformative: they are believed to trigger central starbursts \citep[][]{Heckman1990,Hopkins2008mergerRemnants,Perez2011}, accelerate the feeding of gas to supermassive black holes \citep{DiMatteo2005,Hopkins2008framework}, and can transform a galaxy's morphology \citep{Barnes1996}. Much observational work has been done verifying qualitative predictions of simulations over a range of redshifts \citep[e.g.][to name a few]{Kennicutt1987,Barton2000,Conselice2003,Koss2010,Xu2012,Cotini2013,Ellison2019}. Despite extensive study, a detailed and fully quantified picture of the merger process and its impacts on various galaxy properties remains a challenge -- a carefully matched control sample is needed to separate effects of various parts of the merger process from intrinsic trends in galaxy populations \citep{Perez2009,Ellison2013,Bickley2022}.

Terminology related to mergers varies and can easily lead to confusion. For consistency and clarity, we describe the stages of the merger sequence with our preferred terminology as follows. Galaxies first orbit each other as a pair that becomes closer (on average) with time due to dynamical friction -- we refer to this as the ``inspiral'' phase. As the pair becomes even closer, it may appear as a single disturbed galaxy but with a double nucleus. We consider a pair to have coalesced when there is a single nucleus. Galaxies that have coalesced but can still be identified morphologically as a merger product (from disturbed or tidal features) are referred to in this work as \emph{post-coalescence mergers}.

Large systematic galaxy surveys, in particular the Sloan Digital Sky Survey (SDSS) of approximately one million nearby galaxies, have enabled much more detailed statistical study of mergers via close pairs. Studies have found bluer bulge colours \citep{Ellison2010,Patton2011,Lambas2012}, enhanced star formation rates \citep{Nikolic2004, Alonso2006, Li2008, Ellison2013, Scudder2012, Patton2013,Lackner2014,Pan2019}, a modest reduction in metallicity \citep[e.g.][]{Scudder2012,Thorp2019}, enhanced \HI\ gas \citep[e.g.][]{Scudder2015,Dutta2018,Ellison2018}, increased AGN activity/fraction \citep[e.g.][]{Ellison2011,Lackner2014,Weston2017}, etc. In particular, modest but significant enhancement of star formation is present in galaxy pairs that have a separation $r_p < 150 h_{70}^{-1}$~kpc \citep{Patton2013}, with star formation rates matching control galaxies beyond this ($150 h_{70}^{-1} \mathrm{kpc} < r_p < 1 \mathrm{Mpc}$), indicating enhanced star formation on the order of a Gyr or more prior to merging \citep[][]{Kitzbichler2008,Jiang2014}.

%Part of the challenge historically has lay with the age-burst degeneracy, where weaker but younger starbursts have a similar effect on appearance as stronger but older starbursts \citep{Leonardi1996,Liu1996}. Using spectral features over a range of wavelengths, e.g. by fitting spectral energy distributions (SEDs) using stellar population synthesis templates can help break this degeneracy \citep[][]{Barger1996,Shioya2002,Falkenberg2009,Du2010,Bergvall2016,French2018}. 

While much of the observational merger literature has focused on close pairs or pre-coalescence mergers with two visible nuclei, little work has been done on post-coalescence mergers. As a result, the amount of stellar mass formed during the final burst remains highly uncertain. A major difficulty has been identifying a large sample of post-coalescence mergers in a consistent way. 

Post-starburst galaxies (PSBs), quenched galaxies with a significant population of type-A stars, indicative of burst of star formation in the last $\sim 1$~Gyr \citep[][]{Gonzalez1999}, are often assumed to be mostly post-coalescence merger galaxies, at least at low redshifts, where starbursts should otherwise be uncommon. Observationally, $50-90$ per cent of post-starbursts feature tidal features or disturbed morphologies \citep[e.g.][]{Pawlik2016,Sazonova2021}. \citet{Ellison2022} find a 30--60x excess of PSBs in post-coalescence mergers (but not for close pairs), lending further support to this connection, but make it clear that less than a majority of post-coalescence mergers are PSBs. By selecting PSBs based on the presence of a strong burst, they may not be representative of the post-coalescence merger population as a whole. 

Few attempts have been made to observationally quantify the amount of stellar mass formed in galaxy mergers. Samples of PSBs find large stellar mass burst fractions of $\sim 0.30$ \citep[e.g.][]{French2018}. \citet{Hopkins2008mergerRemnants} found a stellar mass burst fraction of $\sim 0.25$ by fitting excess central light in a sample of $\sim 50$ morphologically-identified gas-rich post-coalescence merger candidates. Very recently, \citet[]{Yoon2023} use stellar ages for a small sample of galaxies with any morphologically-identified tidal features and find a burst fraction of up to 7~per cent for their $10.6<\logMstellar<11.1$ galaxies.

The goal of this work is to model the stellar mass burst during merging using galaxy stellar ages. To measure the stellar mass created in the starburst from the coalescence stage of a typical $1<\mu < 10$ merger in a systematic and unbiased way, a sample must ideally have high purity (high fraction of genuine mergers) and highly completeness or representative sample of post-coalescence mergers. By using the machine learning-identified but visually confirmed post-coalescence mergers of \cite{Bickley2022}, we expect to have a highly pure and representative sample of post-coalescence mergers.

The outline of this paper is as follows. In Section~\ref{sec:data-and-sample-selection} we describe the SDSS observational data and sample selections of post-coalescence mergers and controls. Then in Section~\ref{sec:modeling-and-results} we present our core results: observed properties of merger galaxies, particularly stellar ages compared to controls, as well as our star formation history modeling of both the inspiral phase and stellar mass excess from the final burst of star-formation during coalescence. In Section~\ref{sec:discussion}, we discuss the robustness of these results and contrast them with gas mass fractions and works in the literature. We conclude in Section~\ref{sec:conclusions}.

Unless otherwise specified, the following assumptions and conventions are used. Uncertainties are estimated from the 16th-84 percentile interval (equivalent to 1-$\sigma$ for a Gaussian-distributed variable). Logarithms with base 10 ($\log_{10}$) are written simply as `$\log$' throughout this work. A flat $\Lambda$CDM cosmology consistent with the Planck 2015 cosmological parameters \citep{Planck2015CosmoParams} is assumed, namely $H_0=68~{\rm km}~{\rm s}^{-1}~{\rm Mpc}^{-1}$, $\Omega_m=0.31$, and $\Omega_{\Lambda}=0.69$. A \citet{Chabrier2003} initial mass function (IMF) is assumed throughout. `Age' of a galaxy refers specifically to mass-weighted age, expressed as a lookback time. Finally, we define the stellar mass ratio of a pair of galaxies as $\mu \equiv M_{\star,1}/M_{\star,2}$ (primary to secondary stellar mass ratio) and use this throughout.

%%%%%%%%%%%%%%%%%%%%%%%%%%%%%%%%%%%%%%%%%%%%%%%%%%
%%%%%%%%%%%%%%%%%%%%%%%%%%%%%%%%%%%%%%%%%%%%%%%%%%

\section{Data and sample selection} \label{sec:data-and-sample-selection}

%%% %%% %%% %%% %%% %%% %%% %%% %%%
\subsection{Observational data: SDSS} \label{SDSS_dataset_description}

\subsubsection{Stellar masses and ages}
We use stellar masses and mass-weighted ages (hereafter simply `ages') of the value-added catalogue \citet{Comparat2017}, who performed full spectral fitting of galaxy properties from the Sloan Digital Sky Survey (SDSS) Data Release 14 \citep{Abolfathi2018} using \textsc{FIREFLY} \citep{Wilkinson2017FIREFLY}. The SDSS data are limited to a Petrosian r-band magnitude $m_r <17.77$~MAG. We only include objects whose spectra \citet{Comparat2017} classified as a `GALAXY'. Note that this excludes objects classified as a `QSO' -- quasi-stellar objects. We compare their stellar masses with those of \citet{Mendel2014} and confirm that an offset of $+0.2$~dex relative to the stellar masses of \citet{Mendel2014} do not affect our results. In particular, we use the fits of \citet{Comparat2017}, which were done using the M11 stellar population models of \citet{MarastonStromback2011}, a Chabrier IMF \citep{Chabrier2003}, and the \textsc{MILES} stellar library \citep{SanchezBlazquez2006,FalconBarroso2011,Beifiori2011} for the mass-weighted ages (`CHABRIER\_MILES\_age\_massW'), stellar masses (`CHABRIER\_MILES\_stellar\_mass'), and SFRs from \citep{Brinchmann2004,Salim2007}.

%%% %%% %%% %%% %%% %%% %%% %%% %%%
\subsubsection{Post-coalescence merger sample}

We choose the visually-confirmed sample, confirmed by full consensus of three of their co-authors, because expert visual classifications have long been the preferred way of identifying and confirming post-coalescence mergers. In this sample, only morphologically-disturbed galaxies that appear to be coalesced, i.e. that do not have a double nuclei, are included. An overview of how the visually-confirmed sample was generated follows (for full details see \citealt{Bickley2021} for their Convolutional Neural Network architecture and \citealt{Bickley2022} for how the sample was generated). 

They use a Convolution Neural Network to identify an initial sample of post-coalescence merger candidates. For training data for the neural network, they convert IllustrisTNG cosmological magnetohydrodynamical simulation galaxies into mock observations using the observational realism code \textsc{RealSimCFIS}, a customized version of \textsc{RealSIM} \citep{Bottrell2019}. In particular, the training set is composed of post-coalescence merger and non-post-coalescence merger galaxies from the 100-1 (100$^3$~Mpc$^3$) run of the IllustrisTNG simulation \citep{Marinacci2018,Naiman2018,Nelson2018,Pillepich2018,Springel2018,Nelson2019}, with post-coalescence mergers identified as having completed a merger within the most recent simulation snapshot ($\sim$170~Myr temporal resolution). Mergers were required to have stellar mass ratios of $1\leq \mu \leq 10$ in the stellar mass range $10^{10}$--$10^{12}~\Msun$ at $z \leq 1$. Non-mergers in this training set were selected to have not experienced a merger in the last 2~Gyr. The \textsc{RealSimCFIS} code added redshift dimming as well as observational artifacts, including noise, which we note should match the surface brightness limit of the Canada-France Imaging Survey (CFIS), indicated implicitly by some tests which \citet{Bickley2022} performed. \citet{Bickley2022} then used this neural network to identify initial candidates from the deep $r$-band imaging of CFIS \citep[now part of UNIONS\footnote{\url{https://www.skysurvey.cc/}}]{Ibata2017}. The 5000 deg$^2$ CFIS survey, with 3300 deg$^2$ survey overlap with SDSS, has typical seeing of 0.6 arcsec and a surface brightness limit of 28.4 mag/arcsec$^2$ (Jean-Charles Cuillandre, private communication). They only included candidates assessed by the CNN as having a high probability of being a post-coalescence merger, namely a `decision threshold' greater than 0.75 (corresponding to a true positive rate of $\sim 0.75$ and false positive rate of $\sim 0.05$, see fig.~5 of \citealt{Bickley2021}). This threshold was chosen to increase purity (e.g. a threshold of $\sim 0.5$ would result in many more non-mergers than mergers, i.e.\ a very low purity, given that mergers are a small minority population among galaxies) and to result in a feasible candidate sample size for visual inspection. It is this sample that was then followed up by visual confirmation to produce a final post-coalescence merger sample. Of the 2000 galaxies selected by the neural network, 699, or 35~per cent, were independently visually confirmed by all three visual inspectors.

\begin{figure}
	\includegraphics[width=\columnwidth]{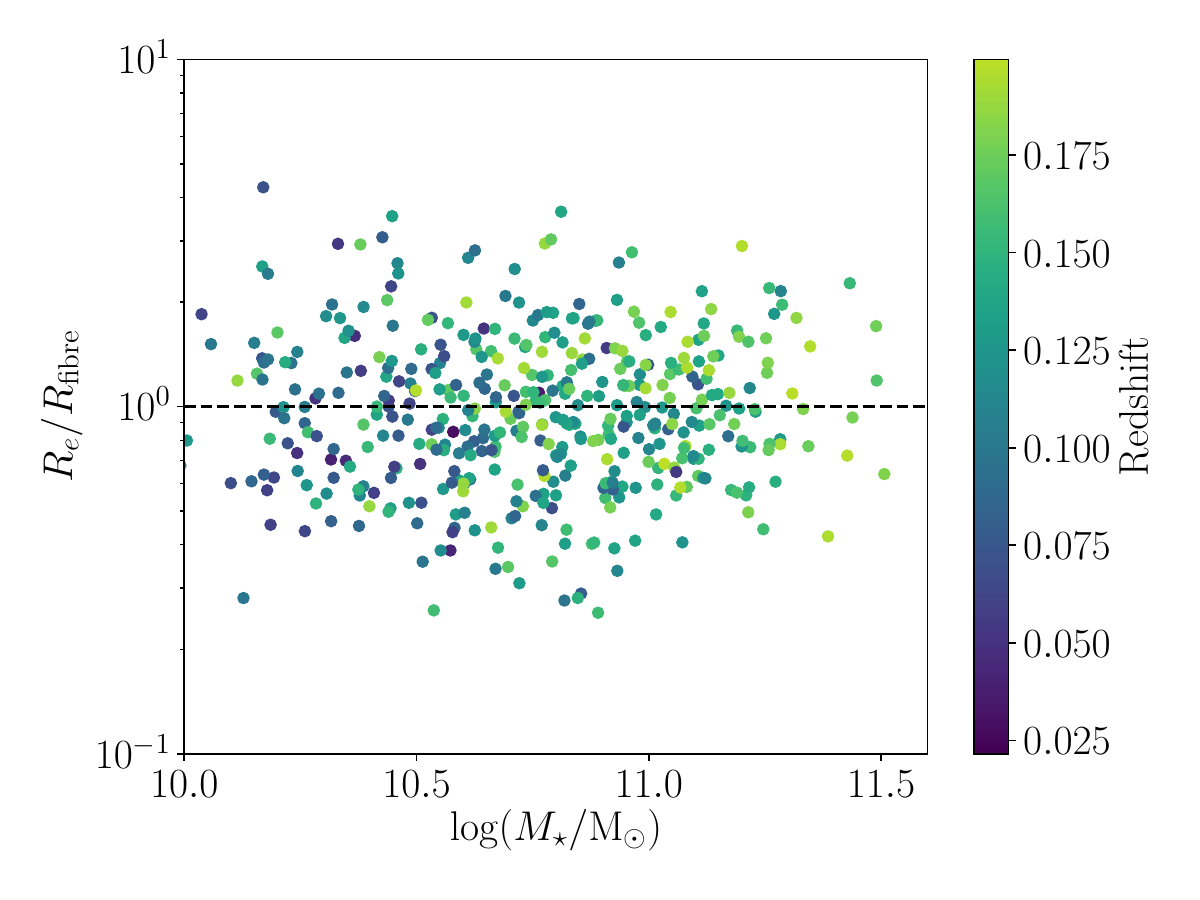}
    \caption{
        Ratio of effective radius to SDSS fibre size versus stellar mass for the post-coalescence merger sample, with redshift denoted by color. The dashed line indicates $R_\mathrm{e} / R_\mathrm{fibre} = 1$; 55~per cent of post-coalescence merger galaxies fall below this line.
    }
    \label{fig:ratio_Re_to_fibre}
\end{figure}

Using the visually-confirmed post-coalescence merger catalogue of \citet{Bickley2022}, we match to the \textsc{FIREFLY} catalogue by \citet{Wilkinson2017FIREFLY},  which has SDSS spectroscopy-derived stellar ages, yielding 560 galaxies out of their total 699 post-coalescence mergers catalogue (some of this loss in galaxies is due to \citet{Wilkinson2017FIREFLY} excluding objects with QSO spectra). We then make a cut to include galaxies with $0.01\leq z \leq 0.2$, $\logMstellar>10$, which leaves us with 445 galaxies. The mean redshift of the merger sample is 0.13 and, since the SDSS fibre aperture is 3 arcsec, the SDSS fibres cover a radius of 3.6~kpc. The ratio of galaxy effective radius to SDSS fibre size, $R_\mathrm{e}/R_\mathrm{fibre}$, using the Sérsic fits from \citet{Simard2011}, versus stellar mass for the  sample is shown in Figure~\ref{fig:ratio_Re_to_fibre}. We can see there is no trend with stellar mass and find that 55~per cent of galaxies in the sample have $R_\mathrm{e}/R_\mathrm{fibre}<1$. We discuss the impact of this effect in our robustness discussion in Section~\ref{sec:robustness}.

% fibre size vs galaxy sizes.ipynb

%%% %%% %%% %%% %%% %%% %%% %%% %%%
\subsubsection{Merger control sample} \label{sec:merger-control-sample}
To construct a control sample for the post-coalescence mergers, we closely follow the control sample selection methodology of \citet{Ellison2013}. Explicitly, the overall collection of all possible control galaxies are those that appear to be isolated, having no spectroscopic companion within $80 h^{-1}$kpc and with a relative velocity of $\Delta v$ within 10~000~km~s$^{-1}$. From these galaxies, we then select matching control galaxies for each post-coalescence merger within a redshift tolerance of $\Delta z = 0.005$, a mass tolerance of $\Delta \log M_{\star}=0.1$~dex, and a normalized local density difference of $\log (1+\delta_5) = 0.1$~dex. Normalized densities, $\delta_5$, are computed relative to the median local environmental density,
\begin{equation*}
    \Sigma_n = \frac{n}{\pi d_n^2}\,,
\end{equation*}
within $\Delta z\pm 0.01$, where $d_n$ is the projected physical distance to the $n$-th nearest neighbour within $\pm 1000$~km~s$^{-1}$. As in \citet{Ellison2013}, we use $n=5$. 

We require that there are at least 5 control galaxies per post-coalescence merger. If there are fewer than this, we increase the tolerance limits (additively) by another $\Delta z=0.005$ in redshift, $\Delta \log M_\star = 0.1$~dex in stellar mass, and $\Delta \log (1+\delta_5) = 0.1$~dex in normalized local density. Only 0.7 per cent of post-coalescence mergers require more than one loop to find more than 5 control galaxies. We end up with $\sim 79~000$ control galaxies total.

%%%%%%%%%%%%%%%%%%%%%%%%%%%%%%%%%%%%%%%%%%%%%%%%%%

\subsection{Observed post-coalescence merger properties}

% "deltaSFR_radialAndVelocityTrends_and_recentMMs_comparison Ellison method.ipynb" -> 3.2 SFR distribution
% saved in: ./figures/fQ_mergers_vs_controls.pdf
\begin{figure}
	\includegraphics[width=\columnwidth]{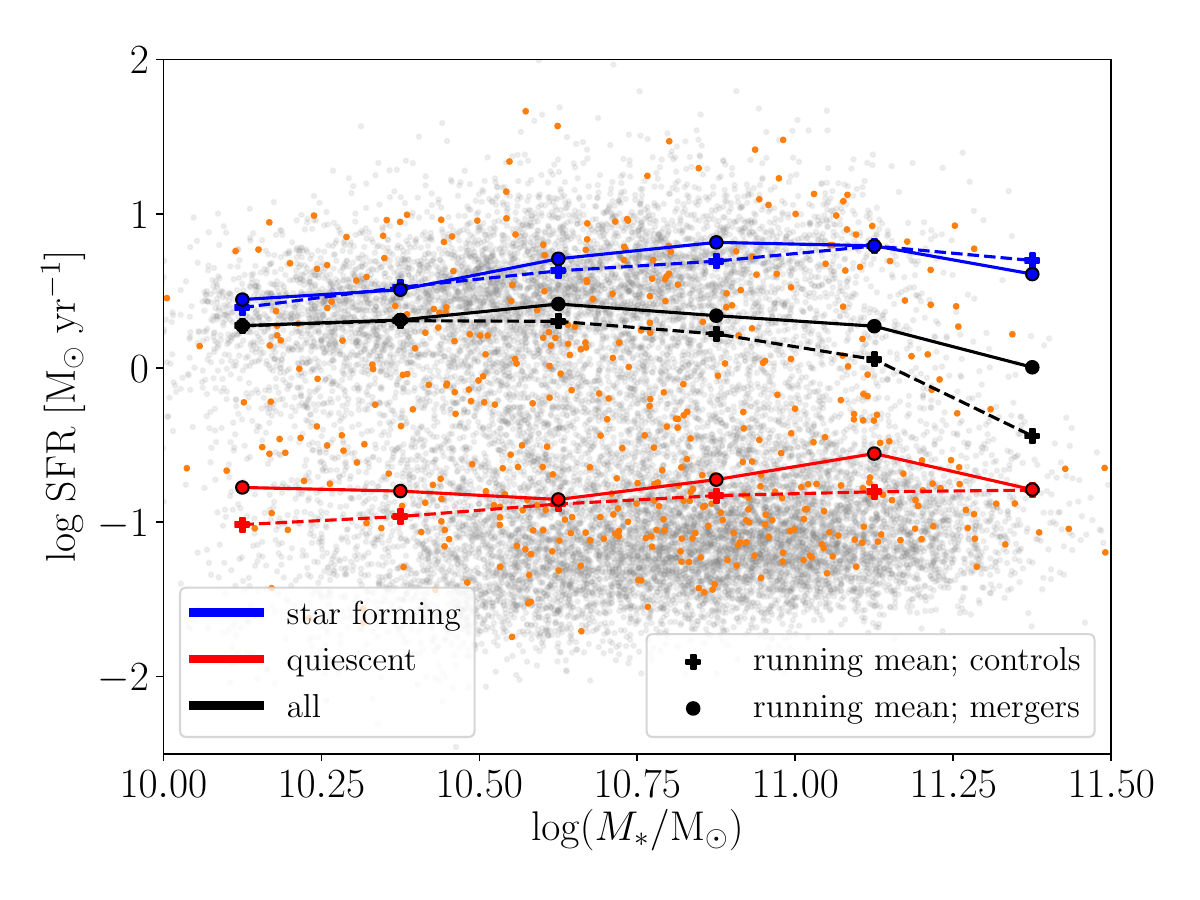}
        \includegraphics[width=0.96\columnwidth]{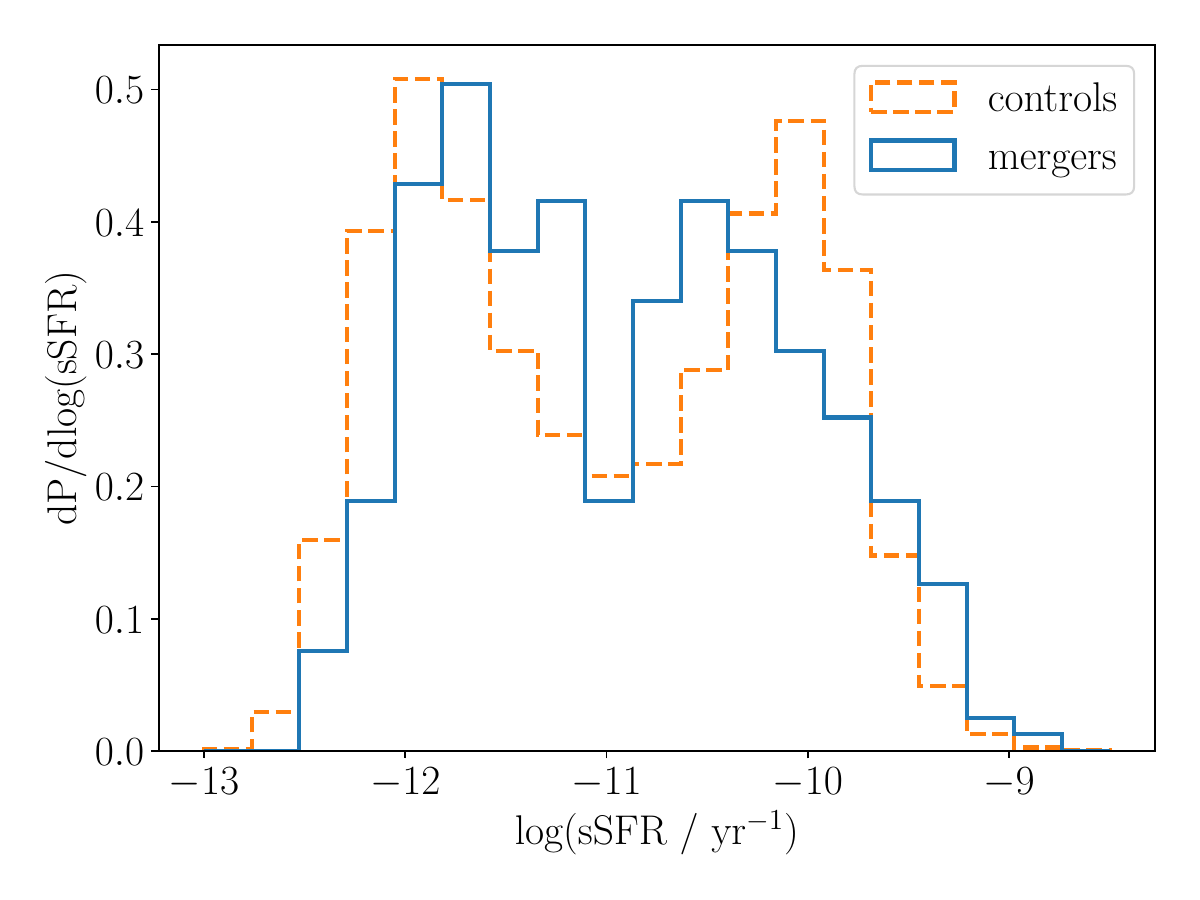}
	\includegraphics[width=\columnwidth]{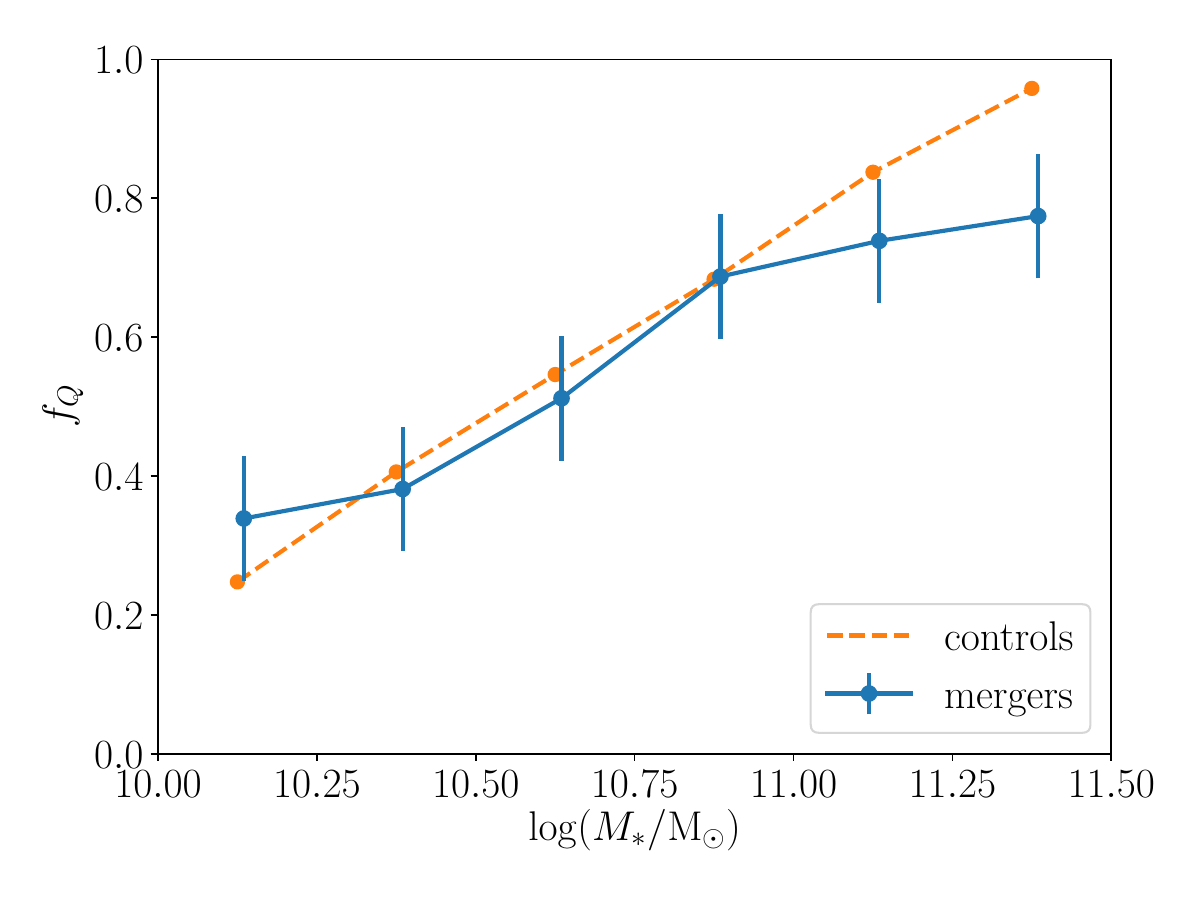}
    \caption{
    \textbf{Top:} Running average SFR for SDSS post-coalescence mergers (circular points connected by solid lines) and controls (`+' symbols connected by dashed lines), broken down by total (black), star forming (blue), and quiescent (red). Each post-coalescence merger's $\log \mathrm{SFR} - M_\star$ value is additionally plotted (small orange points), as is a random subset of the overall SDSS sample (small grey points). \textbf{Middle:} Histograms of the normalized sSFR for post-coalescence mergers (blue solid) and controls (orange dashed).
    \textbf{Bottom:} Quiescent fraction as a function of stellar mass for the post-coalescence merger sample (blue points connected by a solid line) and our control sample (orange points connected with a dashed line). Errorbars on the mergers are the bootstrapped error on the mean $f_Q$; because of the large control sample, errors on the controls' $f_Q$ values are negligible.
    }
    \label{fig:fQ_mergers_vs_controls}
\end{figure}

% plot_Ages_vs_Stellar_Mass.py
% calcs_and_model_scripts/outputs/outputs_OG/plots"
\begin{figure}
	\includegraphics[width=\columnwidth]{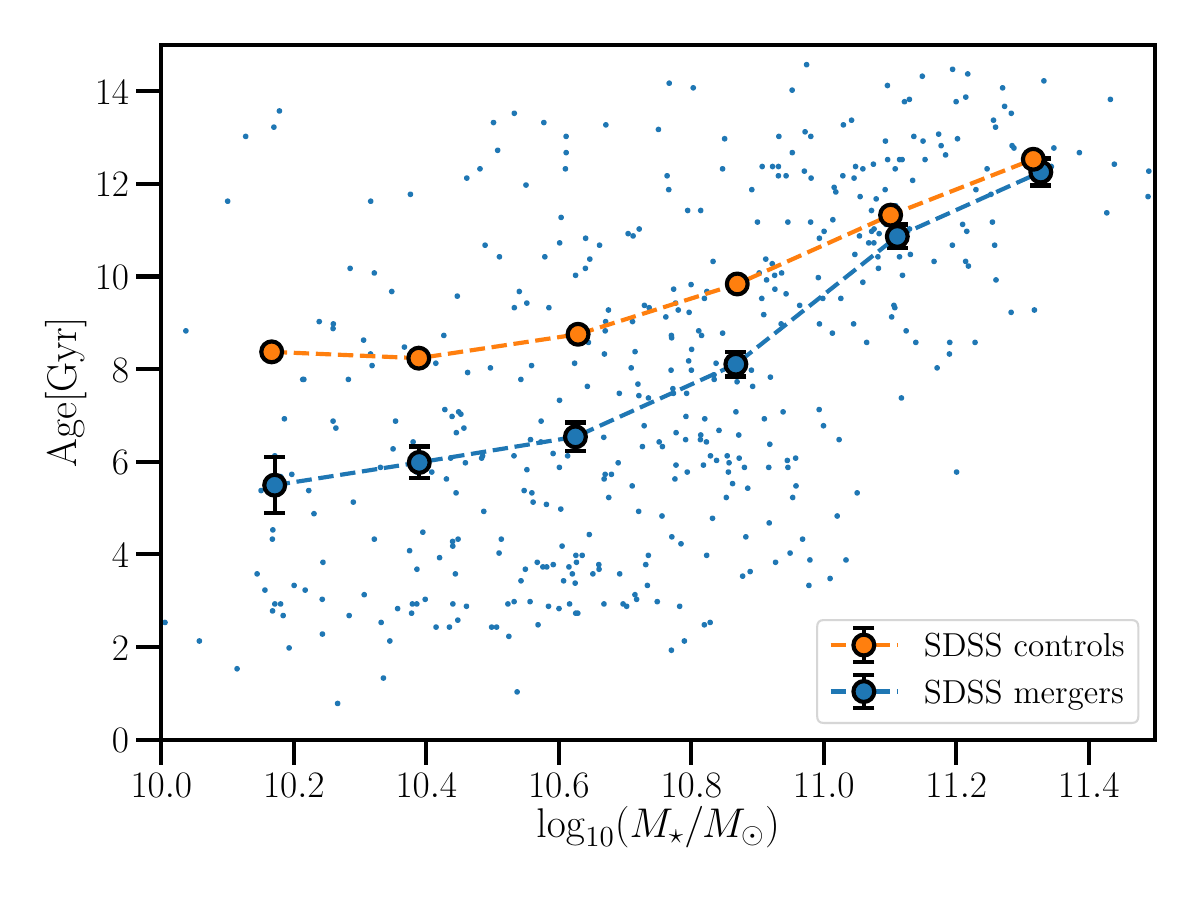}
    \caption{
        Running mean of the SDSS-derived ages as a function of stellar mass for the merger sample of \citet{Bickley2022} (large blue circles) contrasted with our control sample (orange). Post-coalescence mergers are younger than control galaxies by $\sim 2-3$~Gyr for $10<\logMstellar<11$. Error bars shown are the bootstrapped standard error on the mean. All age--$M_\star$ values of the post-coalescence merger sample are shown (small blue points) to give an indication of the scatter in the SDSS ages.
    }
    \label{fig:mergers_Ages_vs_Stellar_Mass}
\end{figure}

We note some observed properties of the post-coalescence merger sample. In the top subplot of Figure~\ref{fig:fQ_mergers_vs_controls}, we see that the log of the running mean SFR of the overall post-coalescence merger sample (black) declines less steeply with stellar mass than for the control sample, but only modestly so. By breaking down the sample into star forming (blue) and quiescent (red), using the sSFR division $\log (\mathrm{sSFR})=-0.4\logMstellar-6.6$, we see the difference is primarily due to the increase in SFR of the quiescent post-merger galaxies relative to the control galaxies. 
%we see that the difference in trend is primarily due to the post-coalescence merger sample having a modestly lower ($<10$~per cent lower) quiescent fraction (bottom subplot of Figure~\ref{fig:fQ_mergers_vs_controls}). 
This is also clear in the distribution of post-coalescence merger sSFRs shown in the middle subplot. Furthermore, the median star forming merger has a lower sSFR than the median star-forming control. In this sense, the mergers populate the ``green valley'' to a greater extent than the controls. Nevertheless, because of a small excess of mergers with much higher star formation rates (log sSFR $\gtrsim -9.8$), we still reproduce the mean $\Delta \mathrm{SFR} \sim 0.27$~dex enhancement in star formation rate\footnote{We note that the mean $\Delta \mathrm{SFR}$ is defined as the mean of $\log \mathrm{SFR}_\mathrm{pair} - \log \mathrm{median}(\mathrm{SFR}_\mathrm{control})$, where $\mathrm{median}(\mathrm{SFR}_\mathrm{control})$ is the median SFR of all matched controls for a given galaxy pair.} for star forming galaxies, as measured in \citet{Bickley2022}, who used the same post-coalescence merger sample. We additionally note that the fraction of quiescent galaxies is consistent between mergers and controls (bottom subplot), except for the very highest stellar mass bin. The reason for the $\sim 20$~per cent lower quiescent fraction in this stellar mass bin presumably has to do with the enhanced proportion of high sSFR galaxies. As this mass bin is outside of our modeled sample, we do not explore it further in this work.

In Figure~\ref{fig:mergers_Ages_vs_Stellar_Mass}, we compare the running mean of the post-coalescence mergers (including both quiescent and star forming galaxies) sample with their corresponding control sample. The well-known age-$\logMstellar$ trend \citep[e.g.][]{NelanSmithHudson2005,GravesFaberSchiavon2007} is apparent for both the post-coalescence mergers and controls sample. The mean age of the post-coalescence merger sample is significantly younger, particularly at lower stellar masses, by up to $\Delta A \sim 3$~Gyr for $\logMstellar \sim 10$. We note that breaking the galaxies down into quiescent or star forming subsets does not change the results, as is expected because \citet{Comparat2017} mask the emission lines (both nebular and AGN) when performing their \textsc{FIREFLY} fitting of the SDSS spectroscopy.

%Since most galaxies with $\logMstellar \gtrsim 11$ are found in the centres of massive dark matter halos, like cluster environments \citep{Yang2008}, and we're only interested in the effects of pair mergers rather than other environmental effects, we limit the modeling of $\Delta A$ to the stellar mass range $10<\logMstellar<11$.
% Properly controlling for the effects of cluster environment is beyond the scope of this work.

%%% %%% %%% %%% %%% %%% %%% %%% %%%
%%%%%%%%%%%%%%%%%%%%%%%%%%%%%%%%%%%%%%%%%%%%%%%%%%

\section{Modeling and Results} \label{sec:modeling-and-results}

Our goal is to determine the burst fraction from $1<\mu<10$ mergers by measuring the fraction of the stellar mass from the starburst that can account for the difference in average age as seen in Figure~\ref{fig:mergers_Ages_vs_Stellar_Mass}. Given the uncertainties on individual observed galaxies' ages, we only use observed average age differences and model average star formation histories. In other words, we are only looking to model the \textit{mean} stellar mass burst fraction and the uncertainty on this mean, and will not be modeling the uncertainty for each galaxy in the distribution, which are very difficult to properly characterise. Note that for $\logMstellar>11$, there is no significant age difference, so we restrict the analysis to $10<\logMstellar<11$. 

We carry out our modeling by adding a recent burst of star formation to the SFR of the pre-merger progenitors, which we model with a parametric SFH, as described in Section~\ref{sec:model-control-SFHs}. We will also need to model the additional star formation during the inspiral phase and so we use published close pair SFR enhancement ratios as a function of pair separation, integrated over inspiral timescales derived from these separations; we present this aspect of the modeling and results in Section~\ref{sec:model-inspiral}. Finally, in Section~\ref{sec:model-starburst}, we present our modeled stellar age results and best-fitting stellar mass burst fraction.

%%%%%%%%%%%%%%%%%%%%%%%%%%%%%%%%%%%%%%%%%%%%%%%%%%
\subsection{Control star formation histories} \label{sec:model-control-SFHs}

For a set of control galaxy star formation histories, we use functions which are log-normal in time, which have been shown to be excellent fits for individual galaxies for most star formation histories, with the exception being the small fraction of galaxies suddenly quenched shortly after becoming satellites (\citealt{Diemer2017}; see also \citealt{Gladders2013}). Following \citet{Diemer2017}, we parameterize our control galaxy star formation histories as 
\begin{align}
    \mathrm{SFR}_\mathrm{con}(t) = \frac{B}{\sqrt{2\pi\tau^2}\times t} \exp\bigg(-\frac{(\ln(t)-T_0)^2}{2\tau^2}\bigg)\,,
\end{align} \label{eq:control-SFH}
where $B$, $T_0$, $\tau$ are free parameters. We note that our results are robust to the particular choice of parametric SFH, as we discuss in Section~\ref{sec:robustness}.

We determine $B$ from the total stellar mass of the galaxy we wish to model, i.e. from their equation~2, $M_\mathrm{final}=B\times 10^9 \times f_\mathrm{ret}$, where $f_\mathrm{ret}=0.6$ is the stellar mass retention factor (similar to that assumed by \citealt{Gladders2013} and that found for IllustrisTNG in \citealt{Diemer2017}).

To solve for the other parameters, we assume the peak time-width relation \citet{Diemer2017} find for their Illustris sample, namely $\sigma_\mathrm{SFR}=0.83 t^{3/2}_\mathrm{peak}$ (their equation~7), where $\sigma_\mathrm{SFR}=2t_\mathrm{peak} \sinh{\sqrt{2\ln{2}}\tau}$ is the full width at half maximum in linear time for the SFH. Let $A_\mathrm{con}$ be the mass-weighted age of the control galaxy. By noting that $t_\mathrm{peak}$ is simply the mode of the log-normal distribution, i.e. $t_\mathrm{peak}=\exp (T_0 - \tau^2 / 2 )$, and that $t_\mathrm{now}-A_\mathrm{con}$ is simply the first moment of the distribution, i.e. $t_\mathrm{now}-A_\mathrm{con}=\exp (T_0 + \tau^2 /2)$, we find the following expression that can be solved numerically for $\tau$ using the mass-weighted age as an input:
\begin{align}
    0 = 0.83 (t_\mathrm{now}-A_\mathrm{con})^\frac{1}{2} \exp\bigg(\frac{-3 \tau^2}{4}\bigg) - 2 \sinh (\sqrt{2 \ln{2}}\tau ).
\end{align}
With $\tau$ now in hand, we can then simply solve for $T_0$ using the definition of the first-moment, $t_\mathrm{now}-A_\mathrm{con}=\exp (T_0 + \tau^2 /2)$.

\begin{figure}
	\includegraphics[width=\columnwidth]{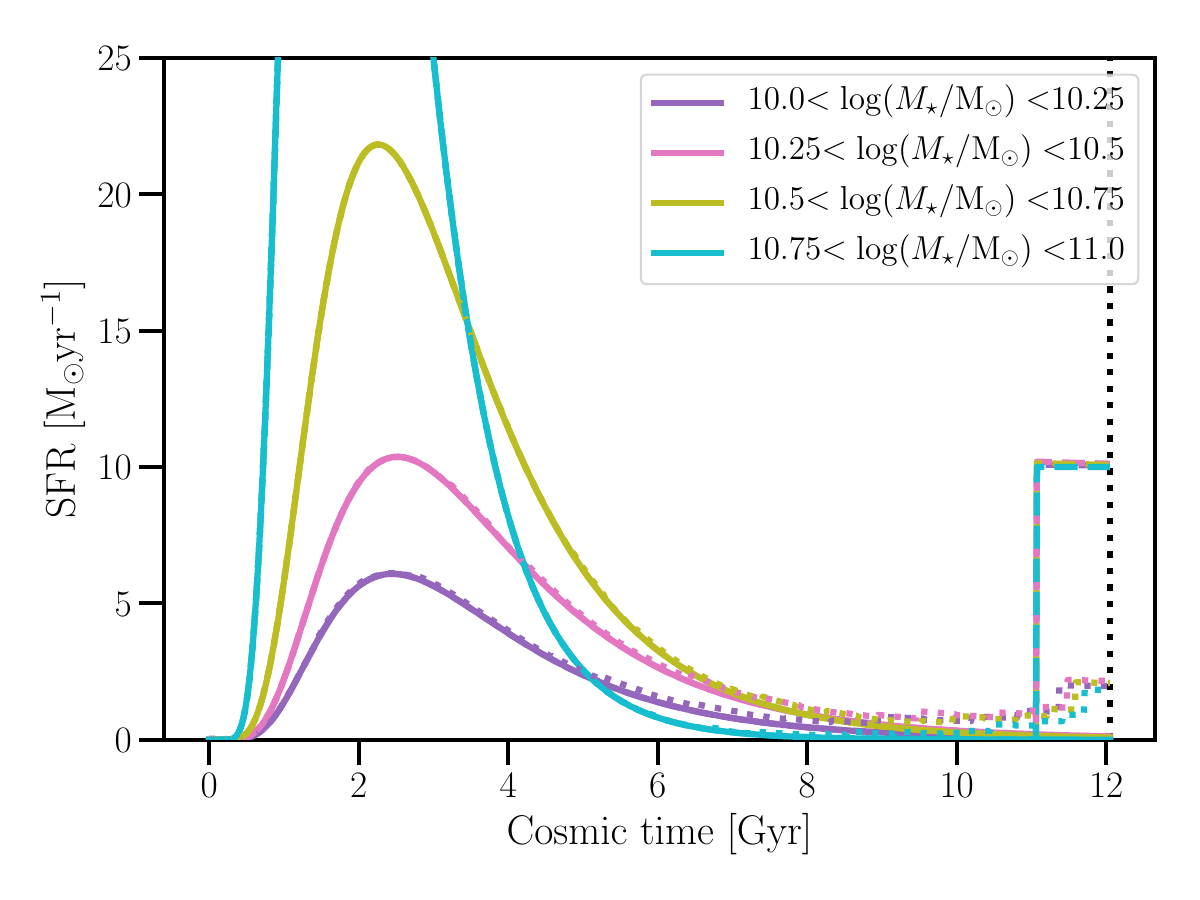}
    \caption{
    Mean log-normal control galaxy star formation histories (solid lines) overlaid with estimated enhanced SFR due to inspiral (dashed lines) binned by stellar mass, as well as an example $\Delta t_\mathrm{burst} = 1$~Gyr additive SFR burst of $\mathrm{SFR}_\mathrm{burst}=10~\mathrm{M}_\odot$yr$^{-1}$. Enhanced SFR from the inspiral phase was calculated using the SFR-$r_p$ galaxy pair results of \citet{Patton2013}, with radii converted to average inspiral timescales using Equation~\ref{eq:tmerge}. Note that star formation histories in this figure are shown up until the average observed redshift, $t_\mathrm{merge}(z=0.13)\sim 12$~Gyr.
    }
    \label{fig:illustration-SFH-plus-inspiral-and-burst}
\end{figure}

We show example log-normal star formation histories for a range of stellar masses, $10<\logMstellar<11$, as the solid smooth curves in Figure~\ref{fig:illustration-SFH-plus-inspiral-and-burst}, using the SDSS control galaxies' age--$M_\star$ relation in Figure~\ref{fig:mergers_Ages_vs_Stellar_Mass} to solve for the SFH parameters for each stellar mass shown.

% calcs_and_model_scripts/illustrations_SFHs_plusInspiralAndBurst_logNormal.py" - "./outputs/outputs_OG/plots" folder

Since we will be modifying these control star formation histories for our merger analysis, rather than use the input stellar mass and age for a given star formation history, we numerically integrate the star formation history. The stellar masses of a galaxy at the average observed redshift of $z=0.13$, is calculated as
\begin{align}
    M_{\star}(t_{\mathrm{now}}) 
    &= \int_0^{t_{\mathrm{now}}} \frac{\diff M_{\star}}{\diff t} \diff t \\
    &= f_{\mathrm{ret}} \int_0^{t_{\mathrm{now}}} \mathrm{SFR}(t) \diff t ,
\end{align}
where $t_\mathrm{now} = 12.05$~Gyr. Similarly, ages are calculated as
\begin{align}
    A_\mathrm{con} &= t_\mathrm{now} - \frac{1}{M_{\star}(t_{\mathrm{now}})} \int_{0}^{t_{\mathrm{now}}} t \Big( \frac{\diff M_{\star}}{\diff t} \Big) \diff t \\
    &= t_\mathrm{now} - \frac{f_{\mathrm{ret}}}{M_{\star}(t_{\mathrm{now}})} \int_{0}^{t_{\mathrm{now}}} t \, \mathrm{SFR}(t) \diff t \,;
\end{align} \label{eq:MWA}
we note it is written this way since it is a mass-weighted age, which is expressed as a lookback time.

%%%%%%%%%%%%%%%%%%%%
\subsection{SFR enhancement during the inspiral phase using close pairs} \label{sec:model-inspiral}

%%% Note: inspiral calculation is done in merger_timescale_explorations.ipynb -> Section 4.0 "Integrated deltaSFR for Patton+2016 pair results; modified Kitzbichler equation"
Previous studies have shown that star formation in the close pair ``inspiral'' phase of the merger is enhanced, as summarised, for example, in table~1 of \cite{Behroozi2015}. To compute the increase in stellar mass during inspiral, we use the empirically-determined relative enhancement in SFR (a ratio to control galaxies) vs $r_p$ relation from figure~1 of \cite{Patton2013}. We convert the projected separation bins to merging timescales by assuming galaxies at a given radius will take some increment of time to fall from one $r_p$ bin to the next, $dt_i = t_{i+1} - t_i$, where $t_i$ is the average infall time for some $r_p$ bin and $t_{i+1}$ is the average infall time for the next farthest $r_p$ bin. The average infall time for a given bin, $r_{p,i}$ is given by $t_i = t_\mathrm{merge} (r_{p,i})$, where
\begin{align*}
    t_\mathrm{merge} (r_p) = 2.2~\mathrm{Gyr} \frac{r_p}{50~\mathrm{kpc}} \Bigg( \frac{\mu}{4} \Bigg) \Bigg( \frac{M_\star}{4\times 10^{10} h^{-1} \Msun} \Bigg)^{-0.3} \Bigg(1+\frac{z}{8} \Bigg),
\end{align*} \label{eq:tmerge}
which is adapted from equation~10 of \cite{Kitzbichler2008} with an extra multiplicative term $\mu$ as found in the fit relation from \cite{Jiang2014}, normalized to $\mu=4$. The extra term is included to correct for \cite{Kitzbichler2008} not examining the dependence of merging time on a pair's mass-ratio (this led to their original expression only matching that of \cite{Jiang2014} for $\mu=4$), which is a significant effect. We note that the mean $\mu$ depends weakly on stellar mass (using only the range $1<\mu<10$), decreasing from $\mu = 3.25$ to $\mu = 2.8$ from $\logMstellar \sim 10$ to $\logMstellar \sim 11$. 

Whether we compute $t_\mathrm{merge}$ using the mean $\mu$ for a given stellar mass bin or whether we calculate the mean $t_\mathrm{merge}$ for a whole distribution of observed pair $\mu$ values does not impact our results. We find that for 2/3 of pairs $t_\mathrm{merge}<1$~Gyr.

The excess stellar mass from the inspiral phase prior to merging is then
\begin{align*}
    \Delta M_\star = f_\mathrm{ret} \sum_i \Delta~\mathrm{SFR}_i \times \mathrm{SFR}_\mathrm{con}(t_i) dt_i ,
\end{align*}
where $\Delta~\mathrm{SFR}_i$ is the ratio in star formation rate between galaxies that are close pairs and their corresponding controls. $\mathrm{SFR}_\mathrm{con}$ is simply the mean SFR of an SDSS galaxy with the control galaxy's stellar mass. The result from this choice is robust as long as most of an inspiral's excess star formation occurs in the last few Gyr, which we will shortly show. We again assume $f_{\mathrm{ret}}=0.6$.

%The fraction of mass lost is $f_{\mathrm{loss}}(t)=0.05 \ln\big( 1 + t/1.4\mathrm{Myr}\big)$ \citep{Behroozi2013} and $t_\mathrm{now}$ is the age of the Universe at $z=0.13$ (i.e. 12.05~Gyr).

In Figure~\ref{fig:illustration-SFH-plus-inspiral-and-burst}, we illustrate the effect of this additional star formation on the overall modeled star formation history due to the inspiral phase (dotted line). Most importantly for this work is the impact on stellar ages from the inspiral phase, which we show with black arrows in the subplots in Figure~\ref{fig:fburst_model_plots}. The enhanced star formation results in younger stellar ages for galaxies, with the effect largest ($\sim 1.5$~Gyr) for the lowest stellar mass bin ($10<\logMstellar< 10.25$), and smallest ($\sim 0.6$~Gyr) for the highest stellar mass bin ($10.75<\logMstellar<11)$. Interestingly, this removes the trend in $\Delta A$ with stellar mass, leaving $\Delta A\sim -1.4$~Gyr at all considered stellar mass bins to be explained by a starburst during coalescence.

%%%%%%%%%%%%%%%%%%%%%%%%%%%%%%%%%%%%%%%%
\subsection{Starburst during coalescence} \label{sec:model-starburst}

We can explain the remaining difference in stellar ages via a simple star-formation burst during coalescence, i.e. upon two galaxies merging. To do this, we model the enhanced star formation during coalescence as an additive burst with a flat SFR, parametrized by the duration of the burst, $\Delta t_\mathrm{burst}$, and the burst fraction for the resulting merged object, $f_\mathrm{burst}\equiv (M_{\star, \mathrm{merger}} - M_{\star,\mathrm{con}})/M_{\star, \mathrm{merger}}$. $M_{\star,\mathrm{con}}$ is the mass of the two merged galaxies without a burst (where `con' is short for 'control'), and $M_{\star, \mathrm{merger}}$ is the final merged object with the burst. We illustrate a simple example $\Delta t_\mathrm{burst}=1$~Gyr long burst of $\mathrm{SFR}=10~\mathrm{M}_\odot \mathrm{yr}^{-1}$ in Figure~\ref{fig:illustration-SFH-plus-inspiral-and-burst}.

This modeling requires an iterative computation to find the input control galaxy's stellar mass. Since our SFH model for control galaxies takes in age as an input, to estimate an age for a trial control galaxy given some stellar mass, we fit the mean age--$M_\mathrm{\star}$ relation for controls with a tight-fitting fourth-order polynomial and flat relation below $x=\logMstellar=10.2$. 
%\begin{equation}
%    A_\mathrm{con}(x)=
%\begin{cases}
%    -2.43+101x-1570x^2+10800x^3-27900x^4, & x>10.2 \\
%    8.36, & x\leq 10.2 .
%\end{cases}
%\end{equation}
We put in a flat trend instead of the polynomial for the lower stellar mass end as there are few post-coalescence mergers (and therefore few matched controls) below $\logMstellar\sim10.2$ and the overall trend for SDSS galaxies is approximately flat below this. We iterate through control masses (the stellar mass of the merged galaxies without the starburst added) until we find the needed control stellar mass to give the desired burst fraction. 

\begin{figure*}
	\includegraphics[width=2\columnwidth]{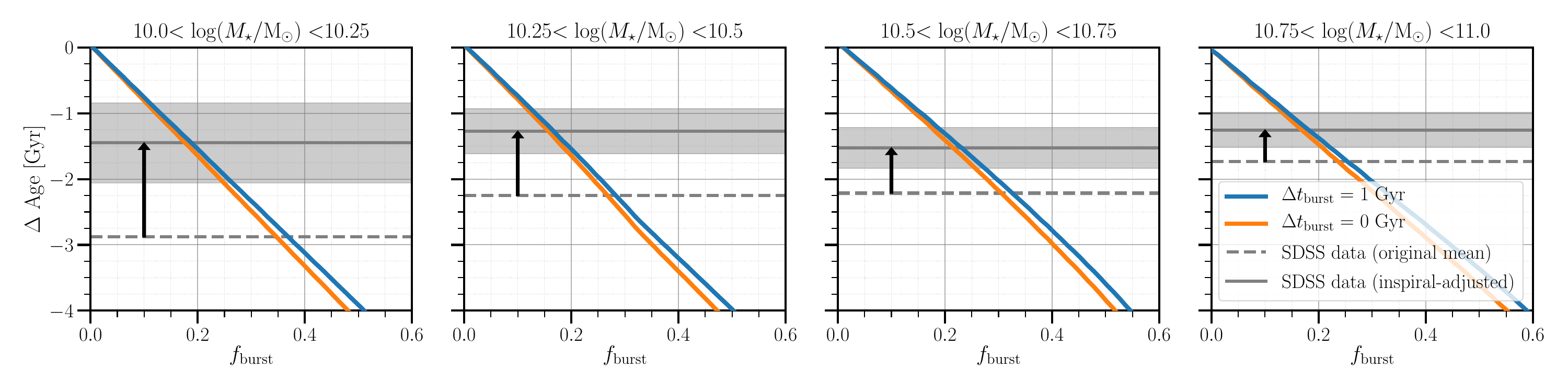}
    \caption{
        Change in modeled merger age, $\Delta A = A_\mathrm{merger}-A_\mathrm{con}$, relative to a control galaxy as a function of burst fraction, $f_\mathrm{burst}=(M_\mathrm{merger}-M_\mathrm{con})/M_\mathrm{merger}$, for four stellar mass bins. On each subplot we show two models: one with an instantaneous burst ($\Delta t_\mathrm{burst}=0$~Gyr) and a model using $\Delta t_\mathrm{burst}=1$~Gyr. In solid grey we plot the mean $\Delta A$ from the SDSS data shifted by the modeled inspiral phase using close pairs data (black arrow indicates the shift), with the shaded region indicating the bootstrapped error on the mean.
    }
    \label{fig:fburst_model_plots}
\end{figure*}

For our model, we plot the change in age, $\Delta A=A_\mathrm{merger}-A_\mathrm{con}$ as a function of $f_\mathrm{burst}$ for our four stellar mass bins in Figure~\ref{fig:fburst_model_plots}. For each stellar mass bin we additionally plot the measured difference in age between the post-coalescence mergers and control sample as a horizontal (grey) band, shifted by the modeled inspiral $\Delta A$ results of the previous section. 

% deltaMWA_vs_fburst_modeling_plots.py
% calcs_and_model_scripts/outputs/outputs_OG/plots"

Earlier works find short star formation bursts. For example \citep{DiMatteo2008} find bursts of up to 500~Myr long at most, with average durations of 200--300~Myr. With this in mind, we show two models with burst durations of $\Delta t_\mathrm{burst}=0$~Gyr (instantaneous; blue line) and $t_\mathrm{burst}=1$~Gyr (long duration extreme case; orange line), which display nearly identical linear declining trends, with a maximum difference between the $\Delta A$ of the two models of $\sim 0.5$~Gyr for high burst fractions, as expected from the difference in age of the two bursts. Additionally, if we perform a simple back-of-the-envelope calculation for an instantaneous burst and assume the control stellar mass is simply the same as the merger, i.e. $f_b = (A_\mathrm{merger} - A_\mathrm{con})/A_\mathrm{merger}$, then we would expect that $\Delta A=-A_\mathrm{con}f_\mathrm{burst}$, which is a very similar linear trend in $\Delta A$ as a function of $f_\mathrm{burst}$ to our plotted $\Delta t_\mathrm{burst}=0$~Gyr model.

\begin{figure}
	\includegraphics[width=\columnwidth]{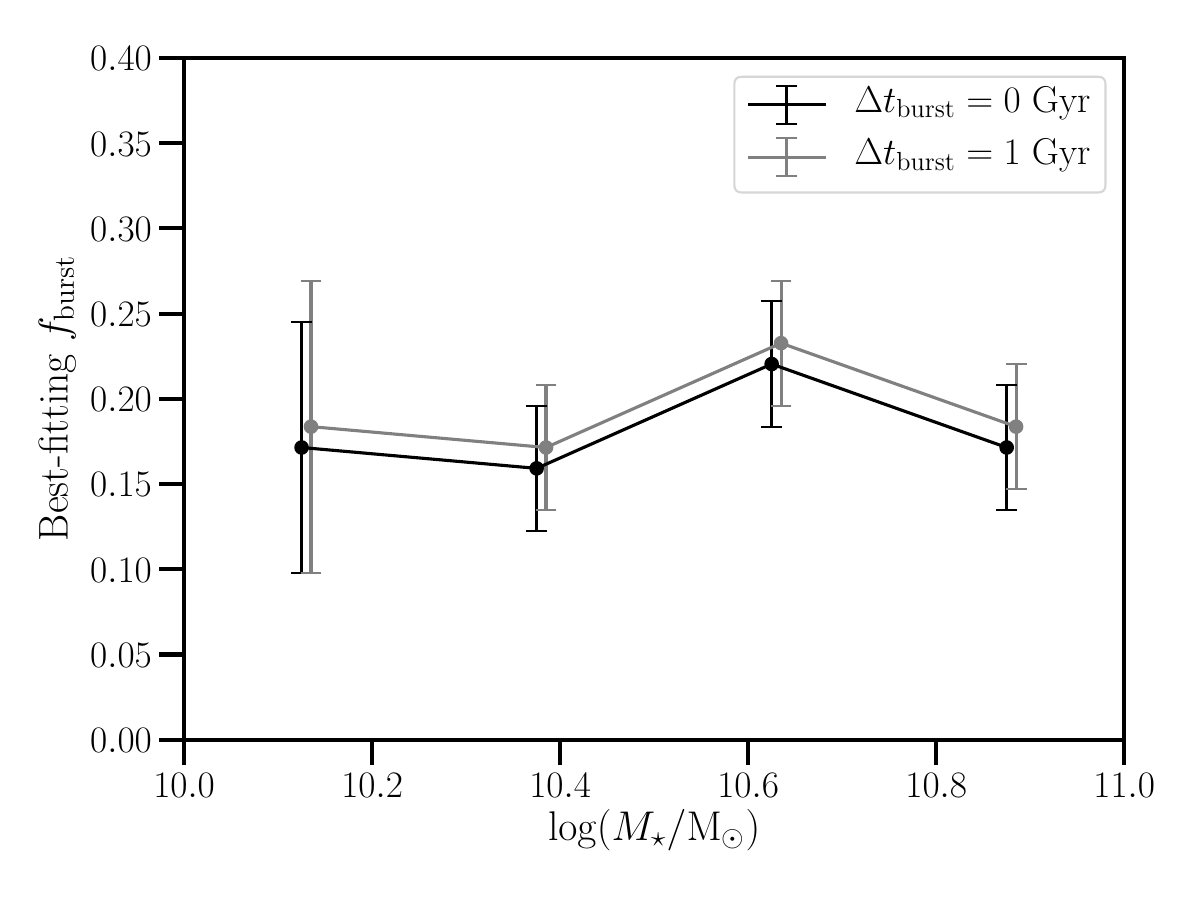}
    \caption{
        Best-fitting burst fraction, $f_\mathrm{burst}=(M_\mathrm{merger}-M_\mathrm{con})/M_\mathrm{merger}$ (for an instantaneous burst), as a function of stellar mass. We show both models from Figure~\ref{fig:fburst_model_plots}: the one with an instantaneous burst ($\Delta t_\mathrm{burst}=0$~Gyr) and a model using $\Delta t_\mathrm{burst}=1$~Gyr.
    }
    \label{fig:fburst_best_fits}
\end{figure}

Assuming the uncertainties on the SDSS ages are normally distributed, we compute the best-fitting $f_\mathrm{burst}$ with uncertainties in each stellar mass bin, which we show in Figure~\ref{fig:fburst_best_fits} for both models. There is a small systematic offset in best-fitting $f_\mathrm{burst}$, such that the $\Delta t_\mathrm{burst}=1$~Gyr is higher by $\sim 0.02$, a difference which is significantly smaller than the uncertainties. We find no trend in $f_\mathrm{burst}$ with stellar mass. The mean stellar mass fraction across our four stellar mass bins is $f_\mathrm{burst}=0.18 \pm 0.02$ ($f_\mathrm{burst}=0.19 \pm 0.03$ for $\Delta t_\mathrm{burst}=1$~Gyr). 

% plot_bestfits_fburst_modeling.py
% calcs_and_model_scripts/outputs/outputs_OG/plots"

%%%%%%%%%%%%%%%%%%%%%%%%%%%%%%%%%%%%%%%%%%%%%%%%%%
%%%%%%%%%%%%%%%%%%%%%%%%%%%%%%%%%%%%%%%%%%%%%%%%%%

\section{Discussion} \label{sec:discussion}

The inspiral phase of galaxy mergers has received significant attention in the literature, in both observational and simulation work, thanks to the ease of observationally identifying close pairs. Since post-coalescence mergers have received relatively little focus in observational work, we focus our discussion primarily on our modeled starburst at the time of coalescence.

%%%%%%%%%%%%%%%%%%%%%%%%%%%%%%%%%%%%%%%%%%%%%%%%%%
\subsection{Robustness of measured burst fraction} \label{sec:robustness}

In this section we explore the robustness of our measured stellar mass burst fraction. In particular, we check, in turn, the effects of the SDSS fibre aperture, the impact of choosing the whole post-coalescence merger sample versus only quenched galaxies, and whether our modeling is sensitive to choice of star formation history.

As noted earlier, the mean redshift of the merger sample is 0.13, for which the SDSS fibres cover a radius of 3.6~kpc. Since the stellar burst is likely mainly in the central regions of the galaxy, most of the burst should fit within the fibre, but some of the galaxy's non-nuclear stellar mass may be cut off. %Assuming that the luminosity distribution is proportional to the stellar mass distribution and assuming that mergers ultimately form ellipticals, one can use the relation between the half-light radius, $R_\mathrm{e}$, and $M_\star$, from figure~8 of \citet{Hyde2009}. 
%The largest effect is for $\logMstellar\sim 11$ galaxies, where $R_\mathrm{e} \sim 3$~kpc, so for those objects the SDSS fibres are only seeing half of the light and hence stellar mass. This suggests that for the most massive galaxies in this paper, the burst fraction may be overestimated by a factor of up to $\sim 2$. For the less massive galaxies the impact of the fibre is negligible: for $\logMstellar\sim 10$, $R_\mathrm{e} \sim 1$~kpc, for which the fibre will capture most of the light \citep[see e.g. figure 1 of][]{Graham2005}.
In Figure~\ref{fig:ratio_Re_to_fibre}, we found that $R_\mathrm{e}/R_\mathrm{fibre}\sim 1$ for an average galaxy in our sample, with little dependence on stellar mass. This implies that for the large majority of galaxies there is no impact on our results, within the margin of error. Additionally, \citet{Yoon2023} studied elliptical galaxies with tidal features (see Section~\ref{sec:discussion-literature-comparison}) and found no difference in the mass fraction formed in the last $2.5$~Gyr, when this was measured within apertures $R < R_\mathrm{e}$ compared with apertures $R > R_\mathrm{e}$. Their results indicate that even if we relax our assumption that the recently-formed stellar mass is centrally concentrated, there is no significant impact on our results from the limited fibre size employed by SDSS.

The merger and control samples have no restrictions on their observed SFRs. For those post-coalescence mergers that are still forming stars at the time at which they are observed, one might expect them to continue doing so for some time after. In that case, the observed burst fraction may underestimate the final burst fraction. This can be checked by using only quiescent galaxies in the analysis. However, restricting to only quiescent galaxies reduces our post-coalescence merger sample from 442 galaxies to 258, with very few galaxies remaining below $\logMstellar<10.5$. For the full stellar mass range we find $f_\mathrm{burst}=0.16\pm 0.04$, with a higher value of $0.24 \pm 0.04$ for $\logMstellar>10.5$, both consistent with our best-fitting value of $f_\mathrm{burst}=0.18\pm 0.02$ for the full sample including both star forming and quiescent galaxies.

We expect our modeling results to be robust to changes in the SFH, as long as the control SFH results in the correct control galaxy stellar mass and age. 
As a test of the robustness of our best-fitting burst fractions to the choice of SFH, we replace our controls' log-normal SFH with a delayed-tau SFH model,
\begin{equation}
    \mathrm{SFH}_\mathrm{con}(t) = 
    \begin{cases}
    B \exp\big(-\frac{t-t_i}{\tau}\big), & t\geq t_i \\
    0, & t < t_i,
\end{cases}
\end{equation}
where $B=M_{\star, \mathrm{con}} / (10^9 f_\mathrm{ret} \tau$) normalizes the star formation history to give the control galaxy's stellar mass. We set $t_i=1$~Gyr as suggested by \citet{Simha2014}, and $\tau$ is chosen such that we reproduce the SDSS controls' age.

Using the delayed-$\tau$ SFH instead of a log-normal SFH results in a lower modeled post-coalescence merger $\Delta A$ by up to 0.09~Gyr, or a lower $f_\mathrm{burst}$ by up to 0.01, for a given value of $f_\mathrm{burst}$ or $\Delta A$, respectively. This effect is greatest for the lower stellar mass bins and negligible for the highest stellar mass bin. From this test it is clear that swapping out our original SFH model with delayed-$\tau$ has no significant impact on our conclusions. We expect this to hold for any reasonable choice of parametric SFH.

\subsection{Dependence on the merger progenitors} \label{sec:wetordry}

Unlike close pairs, with post-coalescence mergers it is impossible to identify the masses, morphological types, and gas fractions of the progenitors of an individual merger. This problem can be studied statistically, however.

The CNN used to identify the post-coalescence merger sample was trained with merger mass ratios, $1 < \mu < 10$, with $\overline{\mu} \sim 3$. What remains uncertain, however, is whether the visual-confirmation step is biased to mass ratios closer to unity. Quantifying this possible effect would require extensive human examination of the mock training set, a rather infeasible task. In any case, our stellar mass burst fraction is not especially sensitive (compared to the uncertainties) to modest changes in the mean $\mu$ (e.g. if $\overline{\mu} \sim 2$ or $\overline{\mu} \sim 4$, rather than $\overline{\mu} \sim 3$).

We note that for a post-coalescence merger galaxy with $\logMstellar\sim 10.2$ with a 20\% burst of stars one might then expect the progenitors to have stellar masses of $\logMstellar\sim 10.0$ and 9.5. These would typically be gas-rich, star-forming galaxies in the ``blue cloud'' and so the merger would be gas-rich or ``wet''. Once the progenitors have stellar masses above $\logMstellar\sim 10.5$, the probability that they will be gas rich drops rapidly. Thus for post-merger remnants with $\logMstellar \gtrsim 10.7$, we expect the burst fraction to drop rapidly as these mergers become ``dry''. Indeed, as shown in Figure~\ref{fig:mergers_Ages_vs_Stellar_Mass}, for $\logMstellar > 11$, the age difference between mergers and controls is consistent with zero (and hence so is the burst fraction).

%%%%%%%%%%%%%%%%%%%%%%%%%%%%%%%%%%%%%%%%%%%%%%%%%%
\subsection{Comparison to other burst fractions in the literature}\label{sec:discussion-literature-comparison}

Most prior work modeling the stellar mass created in a merger starburst have been in complex hydrodynamical simulations or semi-analytic models, but a few works derive stellar mass burst fractions directly from observations of post-coalescence mergers.

In particular, \citet{Yoon2023} use FIREFLY to fit star formation histories to MaNGA integral field unit spectroscopic data of 193 early-type galaxies (ETGs; visually identified using the SDSS $g$, $r$, and $i$ band imaging), 44 of which ($23$~per cent) display tidal features \citep[see][for sample selection]{Yoon2020}. Tidal features were identified visually in the coadded imaging of the Stripe 82 region of SDSS, which is $\sim 2$~mag.\ deeper than the rest of SDSS (UNIONS is in turn $\sim 0.3$~mag.\ deeper than Stripe 82). They found that ETGs with tidal tail features have 
%shallower metallicity gradients and also 
younger stellar ages than those without by $1-2$~Gyr for our stellar mass range, shorter than the $2-3$~Gyr we see between post-coalescence mergers and controls.

They find the fraction of stellar mass formed in ETGs with tidal features in the past $2.5$~Gyr is $2\pm 1$ and $7\pm 3$~percent higher than those without, at $10<\logMstellar\leq 10.6$ and $10.6\leq \logMstellar\leq 11.1$, respectively. The fraction of stellar mass which was formed in the previous $2.5$~Gy (which includes inspiral) that they find for ETGs with tidal features is only $13\pm 3$~per cent and $5\pm 1$~per cent, for $10<\logMstellar\leq 10.6$ and $10.6\leq \logMstellar\leq 11.1$, respectively. They find the same fractions when measuring either within $R < R_\mathrm{e}$ and within $R > R_\mathrm{e}$, meaning we should be able to compare with their result directly. As well, we note that our result is robust to whether we use our entire sample or only quenched galaxies.

Their stellar mass burst fraction is substantially lower than the burst fractions that we find, possibly due to a few important factors. Selecting only for ETGs may bias towards galaxies that have less cold gas than our sample which may contain a proportion of disc galaxies \citep[discs can re-form post-merger, e.g.][]{Hopkins2008mergerRemnants}. However, we find that our results do not change when we consider only quenched galaxies. Including faint tidal features in their selection may also bias the sample towards minor mergers, which are much more common than major mergers. Finally, whereas the sample we use is primarily of recent post-coalescence mergers, \citet{Yoon2023}  selected on generic tidal features, and these may include all stages of the merger process from small satellites stretched out into a tidal stream after first pericentre to $\sim 3$ Gyr after coalescence \citep[although see also][who suggest galaxies selected by tidal features alone are mostly post-coalescence mergers]{Desmons2023}. 
%We note that compared to our inspiral-period excess stellar mass estimate, their result is lower than our $\sim 7$~per cent for $\logMstellar\leq 10.6$ galaxies and higher than our $\sim 1$~per cent for $10.6\leq \logMstellar\leq 11.1$ galaxies.

\citet{Hopkins2008mergerRemnants} studied morphologically-identified gas-rich merger candidates from \citet{Rothberg2004} and fitted surface brightness profiles to quantify the excess central light created by a recent starburst(s). 40 out of 52 of the galaxies are classified as fully violently relaxed in \citet{Rothberg2004}, although our conclusion from their result is not changed whether we include/exclude those with relaxation classified as `incomplete'. In other words, their sample and result should be representative of recently coalesced gas-rich mergers. They find an average best-fitting excess light fraction of $f_e= 0.25\pm 0.03$ %(bootstrapped error on the mean; standard deviation of the distribution is 0.19). 
for galaxies with $\logMstellar \sim 11$, consistent with our burst fraction of $0.18\pm 0.04$ at this stellar mass (including the inspiral period of $\sim 0.01$). Systematic factors such as overestimation of burst mass to burst light fraction, due to reliance on one photometric band, may have resulted in modest overestimation of the stellar mass burst fraction.

\citet{French2018} perform detailed SED-fitting and modeling of post-starburst galaxies, objects for which a substantial fraction appear to be post-coalescence mergers \citep[e.g.][]{Sazonova2021,Ellison2022, Wilkinson2022}. We note that post-starburst galaxies are only a minority ($\sim 20$~per cent) of our post-coalescence merger sample \citep{Ellison2022}. Therefore, it is perhaps not surprising that they find a much higher mean burst stellar mass fraction of $\sim 0.5$ for $\logMstellar\sim 10$ galaxies (versus our $0.28\pm 0.07$ value including the inspiral period). For more massive galaxies with $\logMstellar > 10.5$ their burst stellar mass fraction of $0.21\pm 0.02$ is consistent with ours. 
%Theoretical work, such as the disc survival model for mergers presented in \citet{Hopkins2009}, derive a prediction for burst fraction as a function of pre-merger cold (atomic and molecular) gas fraction of the form $f_\mathrm{sb}=f_\mathrm{gas} (1-f_\mathrm{gas})$. In simple terms, this expression comes about since the starburst fraction will be proportional to the amount of gas multiplied by the stellar mass in the bar (the bar torques the gas), which is proportional to the total stellar mass in the galaxy's disk, namely $(1-f_\mathrm{gas})$. They confirm their result for multiple mass profiles and against a hydrodynamical simulation \citep[see][for further details]{Hopkins2009,Hopkins2013}. Assuming observed gas fractions for close pair galaxies just before merging (which we discuss in more detail in Section~\ref{sec:discussion-gas}), their model predicts burst fractions of 0.15--0.23 for our stellar mass range, with higher burst fractions for lower stellar mass galaxies. The stellar mass burst fraction is indeed consistent with what we see. Note the exact trend is largely a reflection of the trend in gas fraction with stellar mass.

It is also interesting to compare our results with the predictions from hydrodynamical simulations. The IllustrisTNG hydrodynamical simulation was used to train the CNN used by \citet{Bickley2022} to identify the post-coalescence mergers used in our work. \citet{Hani2020} examined post-coalescence merger galaxies in IllustrisTNG and found a modest $\sim 2$ factor increase in SFR, which quickly declines, resulting in only a small $f_\mathrm{burst}\sim 0.5$ per cent. Their result is in clear contradiction with ours and the papers discussed above, apart from \citet{Yoon2023}. \citet{Moreno2019} used the Feedback in Realistic Environments 2 (FIRE-2) hydrodynamical simulations to study pairs of merging galaxies at a 1~pc resolution. Star formation becomes enhanced around the time of first pericentre, followed by a significant (mostly central) starburst with $\mathrm{SFR}\sim 10~\Msun \mathrm{yr}^{-1}$ beginning at second pericentre, $\lesssim 250$~Myr prior to coalescence. Integrating the excess SFR for their $3\times 10^{10}~\Msun$ and $1.2\times 10^{10}~\Msun$ simulated progenitor galaxies (combined stellar mass of $\logMstellar \sim 10.6$), we find 5~per cent of the post-coalescence merger's final stellar mass comes from the inspiral period (prior to second pericentre) and 8~per cent from the burst. This is consistent with our inspiral period's excess stellar mass fraction, but less than half of our best-fitting stellar mass burst fraction for the starburst.

%%%%%%%%%%%%%%%%%%%%%%%%%%%%%%%%%%%%%%%%%%%%%%%%%%
\subsection{Starburst duration}

% plot_bestfits_fburst_modeling.py
% calcs_and_model_scripts/outputs/outputs_OG/plots"
\begin{figure}
	\includegraphics[width=\columnwidth]{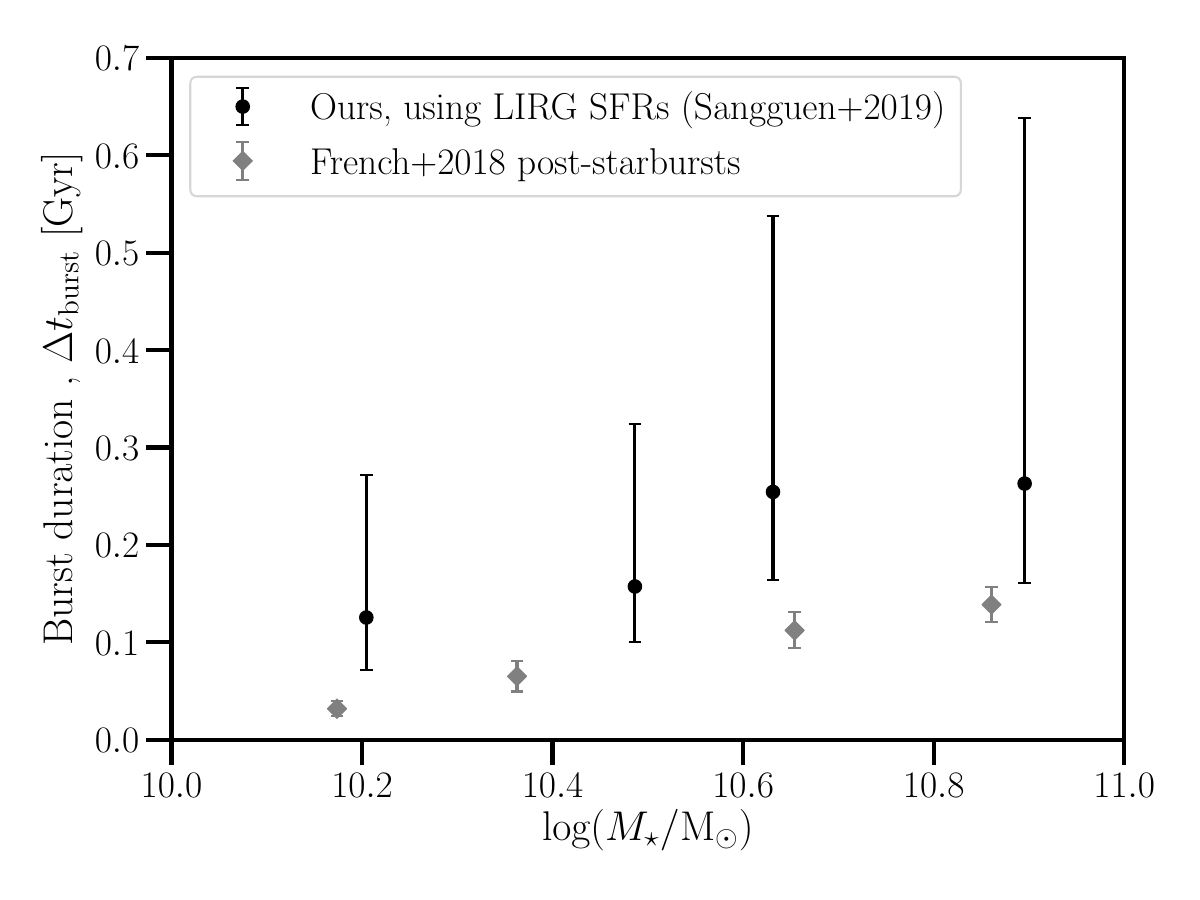}
    \caption{
        Estimated starburst duration, $\Delta t_\mathrm{burst}$, as a function of stellar mass, estimated by assuming $f_\mathrm{burst}$ from Figure~\ref{fig:fburst_best_fits} (black points) and the `late-stage' inspiraling LIRG SFRs from \citet{Shangguan2019} (a presumed population of starburst galaxies).
    }
    \label{fig:fburst_duration_from_LIRGs}
\end{figure}

We can constrain the duration of the starburst during coalescence, if the typical SFR during the starburst is known. Luminous infrared galaxies \citep[LIRGs,][]{Sanders1996} are believed to be starburst galaxies, with $\mathrm{SFR}>10~ \mathrm{M}_\odot$yr$^{-1}$ \citep[similar SFRs to a typical starburst sample,][]{French2018}, usually inferred from IR luminosity \citep[although we note that in principle AGN could be contributing to this, e.g.][]{Iwasawa2011,Petric2011}. As described earlier, the vast majority of low-$z$ starburst/post-starburst galaxies may be due to mergers (see Section~\ref{sec:introduction}). LIRGs/ULIRGs have substantial young stellar populations ($\leq 100$~Myr), and appear to have gone through a period of enhanced star formation prior to their current burst. There is also some correlation with being late-stage inspiraling pairs and especially with coalescing/coalesced merger galaxies \citep[e.g.][]{Gao1997,Rodriguez2010,Stierwalt2013,Larson2016}.

Because of the likelihood at least the majority of LIRGs are due to mergers, we use a sample of 52 late-stage inspiraling LIRGs' SFRs from \citet{Shangguan2019} to measure the typical merger SFR and so constrain the starburst duration. %These LIRGs are a subset of the Great Observatories All-sky LIRG Survey \citep[GOALS,][]{Armus2009}. 
We choose the subset of these morphologically identified by \citet{Stierwalt2013} as having two nuclei in a common envelope. 

We fit a power-law to the $ \mathrm{SFR}$--$M_\star$ trend for this subset of objects (see objects labeled `(d)', e.g. in their figure~5), finding $\mathrm{SFR}= (29 \pm 13) M_\star^{0.53 \pm 0.17}$. Since our modeled burst fraction result is quite insensitive to choice of $\Delta t_\mathrm{burst}$, we can constrain the burst duration 
%using these SFRs by assuming the amount of stellar mass in the burst is 
\begin{align}
    %\Delta M_\star &= %f_\mathrm{ret}\times %\mathrm{SFR_\mathrm{burst}} %\times \Delta t_\mathrm{burst}\\
    %\implies 
    \Delta t_\mathrm{burst} &= \frac{\Delta M_\star}{f_\mathrm{ret} \times \mathrm{SFR_\mathrm{LIRGs}}}.
\end{align}
We show the result of this calculation in Figure~\ref{fig:fburst_duration_from_LIRGs}. We see $\Delta t_\mathrm{burst}\sim 120$~Myr for the lowest stellar mass bin, increasing to $\Delta t_\mathrm{burst}\sim 250$~Myr for the highest stellar mass bins, albeit with large uncertainties from the LIRG SFR--$M_\star$ relation. We note our burst times are similar to the free-fall or violent relaxation time at the outer edge of the disk.
%, since the violent relaxation timescale is $\sim$ the free-fall time \citep{LyndenBell1967}.
% $\sim 400$~Myr for a $\logMstellar\sim 10.75$ Milky Way-mass galaxy ($\logMhalo\sim 12.3$), assuming $R\sim 100$~kpc.

\citet{French2018}, who studied post-starburst galaxies, found an average duration of $103\pm 23$~Myr for $10<\logMstellar<11$ galaxies. Their starburst duration increases with stellar mass, from $30$~Myr for $\logMstellar\sim 10$ to $140$~Myr for $\logMstellar\sim 11$, and is systematically shorter than our estimate, as shown in Figure~\ref{fig:fburst_duration_from_LIRGs}. %Simulations such as FIRE-2 find a final starburst in mergers should start around second pericentre. Since typical coalescence is $\sim 250$~Myr later and \citet{French2018} implicitly select only galaxies that are coalesced, and since they also exclude galaxies with remaining star formation in their disk, their sample is almost certainly ``biased'' towards more rapid and intense starbursts than typical bursts in our post-coalescence merger sample. 
By construction, PSBs are selected to have strong spectral signatures of a burst, whereas the sample we've used is morphologically-selected to have post-coalescence merger features, regardless of whether there was a burst or not. This likely explains the difference in intensity and duration of starburst between the sample used by \citet{French2018} and that used by this work.

\citet{Hani2020} trace their Illustris TNG100-1 $\logMstellar>10$ simulated post-coalescence merger sample forward in time and find that significant enhancements in SFR last for $100 - 250$~Myr post-coalescence (with uncertainty coming from the 162~Myr temporal resolution for the simulation), with a total decay time of $\sim 0.5$~Gyr for this enhancement. This effect was independent of the merger mass ratio. Their timescale is consistent with our estimate. Their SFR burst peaks at a factor only $\sim 2$ higher than their control galaxy, much lower than the factors of 40 -- 100 seen in LIRGs. In the higher resolution FIRE-2 merger simulations, \citet{Moreno2019,Moreno2021} find a longer burst duration than we do: 0.5~Gyr (beginning at second pericentre and therefore finishing 250~Myr after coalescence) for their $\logMstellar\sim 10.6$ simulated merger.

%%%%%%%%%%%%%%%%%%%%%%%%%%%%%%%%%%%%%%%%%%%%%%%%%%
\subsection{Is there enough cold gas to fuel the burst?} \label{sec:discussion-gas}

% calcs_and_model_scripts/outputs/outputs_OG/plots"
% plot_bestfits_fburst_modeling.py (OLD: plot_Mstellar_burst_frac.py)
\begin{figure}
	\includegraphics[width=\columnwidth]{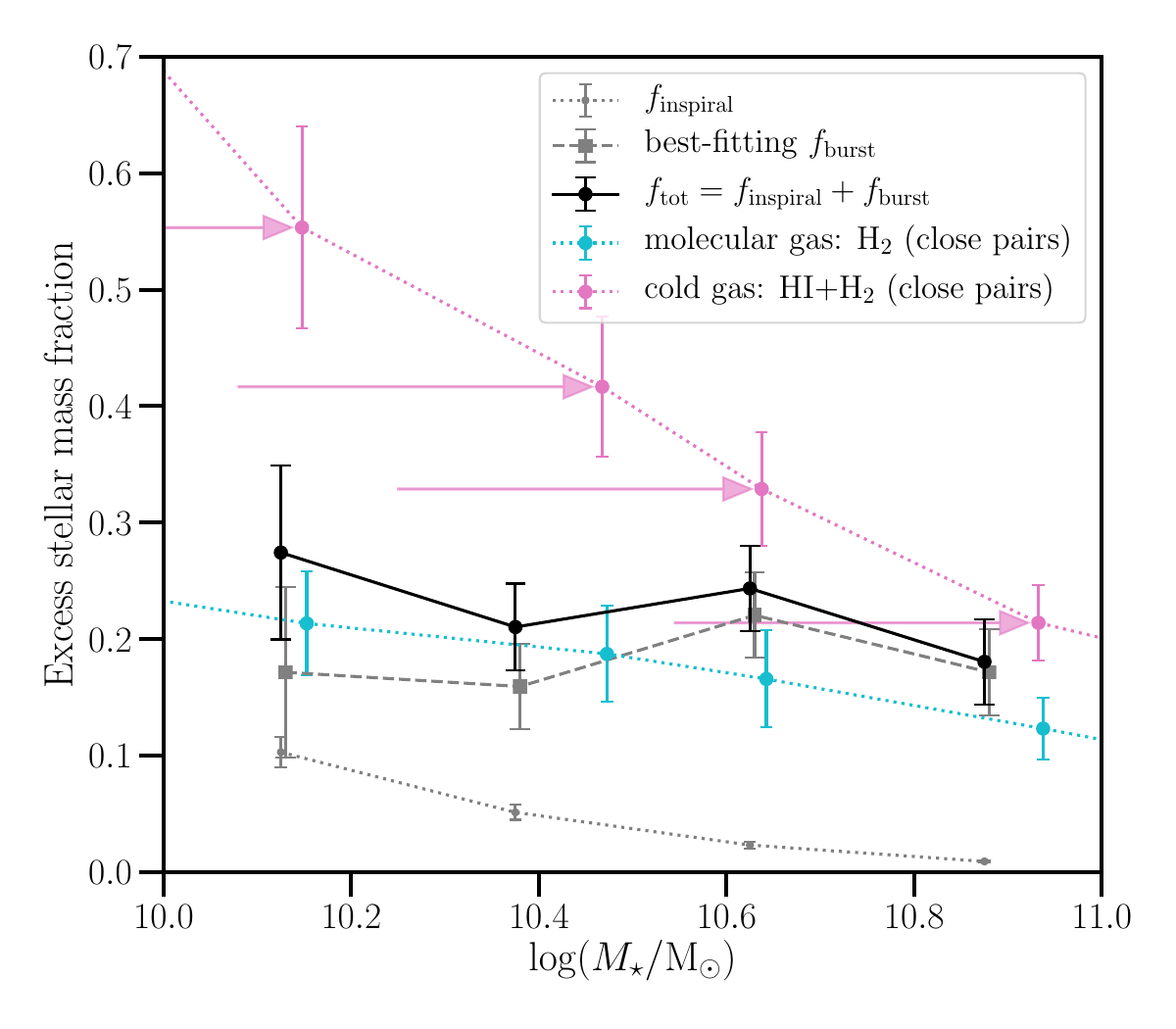}
    \caption{
        Comparison of estimated stellar mass burst fractions from our modeling of the inspiral phase (small grey circles), best-fitting $f_\mathrm{burst}$ (grey squares), and the total excess stellar mass from these two contributions (black points).
        Also shown are molecular (blue diamonds) and total cold gas (pink circles) converted into the equivalent mass fraction of long-lived stars by multiplying by $f_\mathrm{ret}=0.6$. Pink arrows show the stellar mass shift for close pairs gas fractions (so they can be compared to post-coalescence merger quantities), which is applied to both molecular and atomic gas; see text for details.
    }
    \label{fig:stellar_burst_fraction_gas_comparison}
\end{figure}

% calcs_and_model_scripts/outputs/outputs_OG/plots"
% plot_bestfits_fburst_modeling.py
\begin{figure}
	\includegraphics[width=\columnwidth]{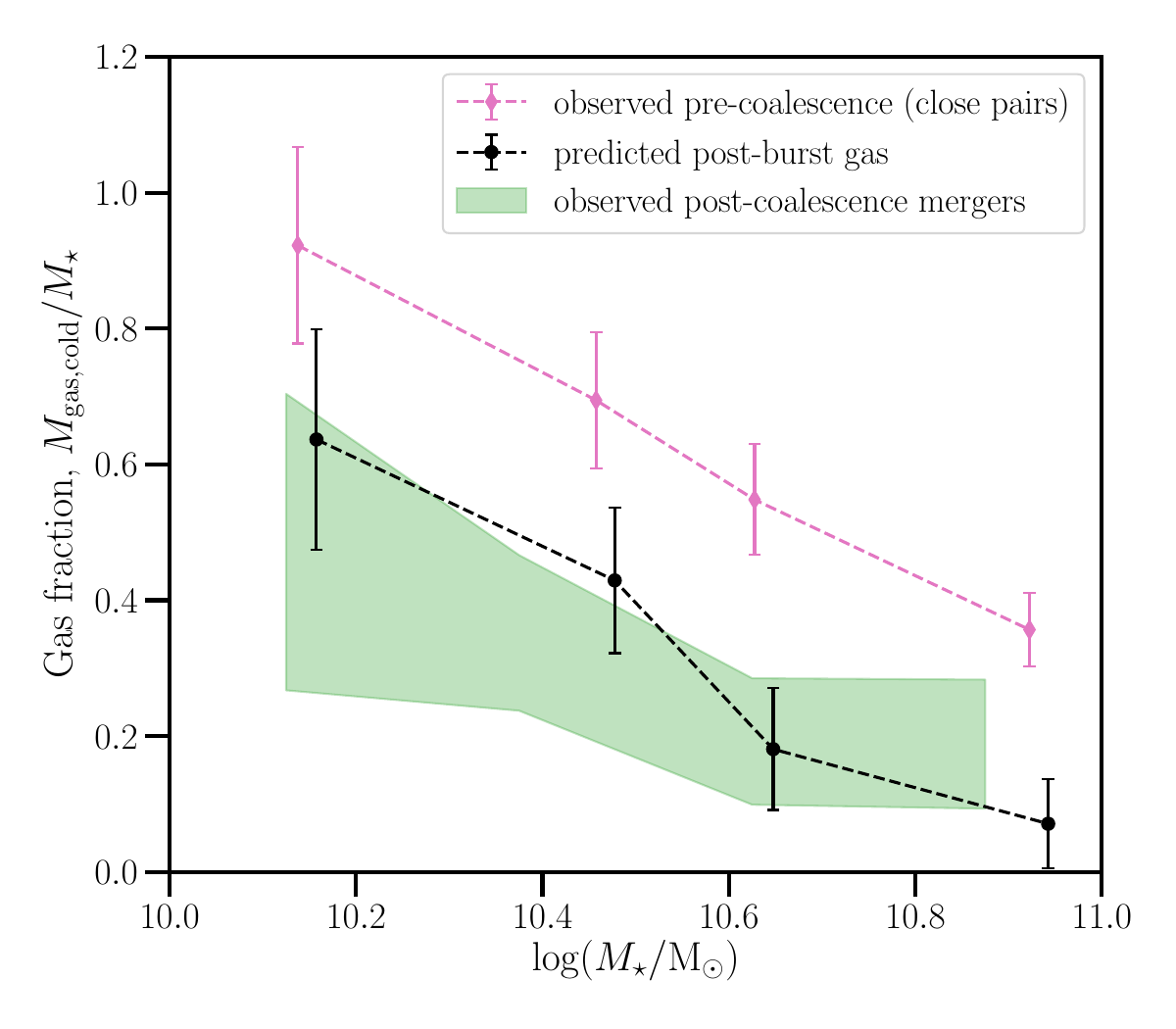}
    \caption{
        Comparison of cold gas before (pink points) and after coalescence (green), as well as our prediction for remaining cold gas after the burst (black points), all as a fraction of the final merger stellar mass. Our prediction is simply the cold gas before minus the gas consumed during star formation (any gas ejected in supernovae is assumed to be hot). For the post-coalescence mergers we use the \HI\ gas masses from \citet{Ellison2018} and $\mathrm{H_2}$ gas masses from various samples of post-starburst galaxies \citep[][see text for details]{French2015,Rowlands2015,Otter2022}.
    }
    \label{fig:cold_gas_before_and_after_coalescence}
\end{figure}

The general picture of cold gas in mergers is as follows. Atomic gas in the galaxy outskirts flows inwards, due to decreased angular momentum from gravitational torques, resulting in rapidly increased cold gas density in the central region. This cold atomic gas condenses into molecular clouds, with collision-induced pressure possibly accelerating the formation of additional molecular gas from atomic (\HI) gas \citep{Moster2011}. This additional molecular gas in the core then leads to intense star formation in the galaxy nucleus \citep{Mihos1996,DiMatteo2008,Renaud2014}. Turbulence induced by gravitational torques during interactions, particularly at the start of coalescence, may also lead to gas fragmentation, forming massive and dense molecular clouds, fueling the intense star formation of a starburst \citep[see e.g.][]{Teyssier2010,Bournaud2011}. 

Such a picture is seen in detailed hydrodynamical simulations. In the FIRE-2 simulations, \citet[]{Moreno2021} find a $0.8$--$1$~dex enhancement of cool atomic and $1.1$--$1.2$~dex enhancement of cold-dense molecular gas mass in the central regions ($r<1$~kpc) at time of second pericentre, coming from the galaxy outskirts ($r=1$--10~kpc). This excess gas in the central region rapidly declines, back to the baseline of the two galaxies evolving in isolation, during the duration of the starburst ($\sim 0.5$~Gyr).

A key question is then whether there is enough cold gas fuel available to form the excess mass of stars formed, namely $\sim 0.2 M_{\star,\mathrm{merger}}$, and if so, how much gas remains after the burst? To examine this question, we look at cold gas measurements for pairs of galaxies to estimate the amount of cold gas available to form stars at the start of coalescence.

We take cold gas fraction measurements of \HI\ and H$_2$ from the XCOLD GASS survey \citep[as presented in][]{Saintonge2017}, a systematic survey of $0.01<z<0.05$ galaxies selected from SDSS to be representative down to $10^9 \mathrm{M}_\odot$ in stellar mass. 
%They employ \HI\ Arecibo observations and derive H$_2$ molecular gas masses via IRAM 30m telescope measurements of the CO (1-0) line and also a CO (2-1) / CO (1-0) excitation correction (using IRAM and APEX data). We note their $\mathrm{H_2}$ gas masses are corrected by a multiplicative factor of 1.36 to account for He and metals.
We then adjust these gas masses by the relative enhancements in the proportions of the cold gas components for close pairs. In particular, we note that \HI\ does not appear significantly different from controls in recent samples of several dozen pair galaxies, with some debate as to the exact impact \citep[][]{Ellison2015,Yu2022}. For pre-coalescence \HI\ gas masses, we simply use the unmodified \citet{Saintonge2017} \HI\ masses. For $\mathrm{H_2}$, we take the $\mathrm{H_2}$ gas mass enhancements, relative to controls, of very close pairs from \citet{Pan2018} (most have stellar mass ratios $1 \leq \mu \leq 10$)
%\footnote{For reference, we note that the radius of a Milky Way-sized galaxy is $\sim 15$~kpc. We also note for the careful reader that their line-of-sight velocity cut for close pairs is $\Delta V<500~\mathrm{km s}^{-1}$.}
and we multiply their figure~8 value (projected separation $r_p<30$~kpc) by the ratio of gas masses for $10<\logMstellar<11$ galaxies to their whole sample, to find a mean molecular gas fraction ($M_{\mathrm{H_2}}/M_\star$) enhancement of $0.62\pm 0.07$ dex \citep[see also][]{Casasola2004,Violino2018}. Such an enhancement increases the $\mathrm{H_2}$ mass fraction of the cold gas from $M_{\mathrm{H_2}}/(M_{\HI} + M_{\mathrm{H_2}})=0.08\textnormal{ -- }0.11$ for the general field sample of \citet{Saintonge2017} to $M_{\mathrm{H_2}}/(M_{\HI} + M_{\mathrm{H_2}})=0.39\textnormal{ -- }0.58$ for close pairs (with higher gas fractions in these given ranges being for lower stellar mass galaxies). We note their result is consistent with \citet{Lisenfeld2019}, when this latter work is corrected to account for He and metals.
% who find $M_{\mathrm{H_2}}/(M_{\HI} + M_{\mathrm{H_2}})\sim 0.5$ for their overall sampling of close pairs, $r_p<30$~kpc\footnote{Note: they use a rather large line-of-sight velocity cut of $\Delta V<2000~\mathrm{km s}^{-1}$.}, for the stellar mass range $10<\logMstellar<10.7$ (we have corrected their work by a factor of 1.36 to account for He and metals). 
Such a large increase in molecular gas during the inspiral process is also consistent with data from LIRGS that are morphologically-defined as having double nuclei in the pre-coalescence stage of merging \citep{Larson2016}.

The relevant gas fraction for a merger involves not the stellar mass of the final merger product but rather the gas fractions expected from the progenitors and summed together appropriately. The final stellar mass is then the sum of the stellar masses of the progenitors (close pairs) plus the retained stellar mass after the cold gas is converted to stars. To compare gas fractions with the stellar mass burst fraction, we therefore shift the stellar mass bins of the gas fractions by doubling their mass and appropriately adding in the stellar mass burst, i.e. $M_{\star}' = 2M_{\star}/(1 - f_\mathrm{burst})$. We note that whether we assume equal mass mergers or a more realistic e.g. $\mu=3$, the impact on the gas fractions for our stellar mass range is negligible. We show these gas fractions overlaid with our best-fitting $f_\mathrm{burst}$ values in Figure~\ref{fig:stellar_burst_fraction_gas_comparison}. Assuming $f_\mathrm{ret}=0.6$, we find there is just enough molecular gas available prior to coalescence to form the stellar mass burst, but only if either there is $\sim 100$~per cent efficiency in converting existing molecular gas to stars or if more molecular gas is formed during coalescence. Using instead the total cold gas content (molecular and atomic) prior to coalescence, the gas consumption efficiency of $e=(f_\mathrm{burst} M_{\star, \mathrm{merger}}/f_\mathrm{ret}) / M_\mathrm{gas,cold}$ ranges from 30--80~per cent, increasing in efficiency from the lowest to highest stellar mass bin.
%Averaged over all stellar mass bins, in order to produce the stellar mass burst of $f_\mathrm{burst}\sim 0.2 M_{\star,\mathrm{merger}}$ favoured by our best-fitting model, there is just enough cold gas available to form the stellar mass burst that we see if the gas is consumed with $\sim 100\%$ efficiency. 
% We discuss additional cooling that appears to occur during and shortly after ($\lesssim 0.5$~Gyr) coalescence in the following section.

A similar comparison has been performed for post-starburst galaxies. In particular, \citet[see also \citealt{Rowlands2015}]{French2018} find a significant decline in molecular gas to stellar mass fraction with increasing post-burst age, which persists after controlling for fraction of stellar mass produced in the recent burst. Assuming an exponentially declining gas fraction, they find a best-fitting timescale of $117-230$~Myr and best-fitting initial molecular gas fractions of 0.4--0.7 at a post-burst age of zero \citep[lower end of this range is consistent with][]{Pan2018}. % (with the upper value calculated using a fitting routine that assumes there is an additional intrinsic scatter term in their fit relation). 
Based on their fit relation, for a post-starburst about $0.5$~Gyr after the beginning of the burst (their mean found for post-starbursts with SED fitting), a post-starburst has $M_\mathrm{\mathrm{H}_2}/M_\star \sim 0.05$. 
% The rapid depletion rate of 2--150$\Msun$yr$^{-1}$ is much higher than the current SFR, suggesting molecular gas is being expelled or destroyed by AGN-driven outflows.
The difference between their results and ours is likely driven by the difference in sample selection, since as noted earlier, only $\sim 20$~per cent of post-coalescence mergers are post-starbursts \citep[][]{Ellison2022}.

Subtracting the gas consumed in the burst from the total expected cold gas fraction in the merging pair pre-coalescence, we predict a residual post-burst cold gas fraction as shown in Figure~\ref{fig:cold_gas_before_and_after_coalescence} (black points). Specifically, our prediction assumes all gas ejected by supernovae (i.e. the (1-$f_\mathrm{ret})=0.4$ fraction of cold gas not retained as stellar mass) is in the form of hot gas. Is this consistent with observed gas fractions in post-coalescence mergers?

For \HI\ gas, we use reported \HI\ gas enhancements in \citet{Ellison2018} of median atomic gas-to-stellar mass ratios in observed post-coalescence mergers (see their figure~4). They find enhancements, relative to xGASS stellar mass-matched controls, of $\sim 0.2$~dex when including only detections above their adopted threshold level (which we refer to as their ``\HI\ upper estimate'') and $\sim 0.4$~dex (``\HI\ lower estimate'') when including both detections and upper limits on \HI\ gas masses. We show this range, including uncertainties, as the green shaded region in Figure~\ref{fig:cold_gas_before_and_after_coalescence}. For $\mathrm{H_2}$ gas masses, there are no published values for post-coalescence mergers yet in the literature. Instead, we adopt a compilation of post-starburst $\mathrm{H_2}$ gas masses \citep[][]{French2015,Rowlands2015,Otter2022} in the stellar mass range $\logMstellar=[10,11]$. We combine the \HI\ and $\mathrm{H_2}$ gas fractions into a post-coalescence merger cold gas fraction, shown in green on Figure~\ref{fig:cold_gas_before_and_after_coalescence} (with the shaded vertical range including the range in possible \HI\ values as well as uncertainties on the gas fractions). Our predicted cold gas fraction is consistent with this range, lending further credence to our estimated burst to stellar mass fraction.
% In particular, molecular gas is reported in the literature to be maintained at elevated levels similar to close pairs immediately pre-coalescence \citep[][]{Ueda2021}. This would imply that the amount of $\mathrm{H_2}$ gas that condensed into the molecular gas phase, between the pre-merger close pairs phase and post-coalescence, must be $\sim$ equal to the starburst mass that was formed ($\sim 0.2 M_{\star,\mathrm{merger}}$). Although molecular gas remains elevated after fueling a fairly significant starburst, atomic gas does not appear to be depleted by this. 
%($\sim 50$ galaxies; see bottom panel of their figure~4).
%This excess gas remaining is puzzling for two reasons: (1) why does the left-over cold gas not continue to form stars and extend the starburst and (2) where does this excess cold gas come from?

As to why this cold gas remains after the burst, suppression of the burst before gas can be depleted by star formation has been proposed to be due to enhanced turbulence from the merging process, as well as from shocks and/or outflows from star formation and (non-QSO) AGN feedback \citep[e.g.][]{Veilleux2013,Sell2014,Rich2014,Mortazavi2019}. These effects could make the ISM stable against gravitational collapse even if cold gas is abundant \citep[][]{Alatalo2015,Smercina2018,vandeVoort2018}. A sample of resolved molecular gas observations of MaNGA post-starbursts in the recent work of \citet{Otter2022} supports this, they find compact but highly disturbed molecular gas unable to form stars efficiently.

%%%%%%%%%%%%%%%%%%%%%%%%%%%%%%%%%%%%%%%%%%%%%%%%%%
%%%%%%%%%%%%%%%%%%%%%%%%%%%%%%%%%%%%%%%%%%%%%%%%%%
\section{Conclusions} \label{sec:conclusions}

In this work, we used the morphologically selected and visually-confirmed post-coalescence merger catalog of \citet{Bickley2022} combined with available SDSS photometric and spectroscopic data to directly model the stellar mass formed in the starburst during coalescence. 
%The new post-coalescence merger catalogue by \citet{Bickley2022} was constructed to have both high \textit{completeness} across a swath of the sky and also high \textit{purity}, by using a Convolutional Neural Network trained on mock observations from simulations of galaxies mergers to pre-select any candidate post-coalescence merger galaxies from the CFIS survey. 
To fit for the stellar mass burst fraction, we forward model the difference in mean age--$M_\star$ relation between post-coalescence mergers and a control sample, controlling for stellar mass, local density, and redshift. In particular, we model the star formation history of control galaxies in four bins across $10<\logMstellar<11$, the inspiral (pre-merger close pair) star formation enhancement, and the final starburst from coalescence. Our main results and conclusions are as follows.
\begin{enumerate}[leftmargin=*, labelwidth=\widthof{0.}]
    \item Post-coalescence merger galaxies are younger than control galaxies by $1.2-1.5$~Gyr, with a smaller age difference for higher stellar mass galaxies.
    \item We find a mean stellar mass burst fraction of $f_\mathrm{burst}=0.18\pm 0.02$, independent of stellar mass and with only a very weak dependence on burst duration. 
    \item Our burst fraction is consistent with some observationally-derived values, namely \citet{Hopkins2008mergerRemnants} measurement of gas rich post-mergers and the higher stellar mass end of post-starbursts \citep[][]{French2018}. We find a notably higher burst fraction than another recent study using stellar ages, \citet{Yoon2023}, which may be due to differences in sample selection; our sample is likely trained on larger mergers of any morphological type, whereas theirs includes any tidally-disturbed ETG, potentially including more minor mergers or pre-coalescence disturbances.
    \item Compared to simulations, our burst fraction is twice that of the hydrodynamical FIRE-2 simulation and much greater than that of the cosmological hydrodynamical IllustrisTNG simulation, which finds a negligible starburst.
    \item Using the star formation rates of published LIRGs that were morphologically-identified as late-stage inspiraling pairs (i.e. not yet coalesced), we estimate the starburst duration for post-coalescence mergers is $\Delta t_\mathrm{burst}\sim 120$~Myr for $\logMstellar<10.25$ galaxies and increasing to $\Delta t_\mathrm{burst}\sim 250$~Myr for $\logMstellar>10.5$ galaxies. This is longer than found in the literature for observed post-starburst galaxies, consistent with the Illustris TNG100-1 hydrodynamical simulation, and $\sim$~half as long as for the high-resolution FIRE-2 hydrodynamical simulation.
    \item We find there is enough molecular gas present in close pairs to fuel the starburst that we measure, assuming a high efficiency in converting molecular gas into stars. Assuming both molecular and atomic gas are available as fuel for star formation during the burst (consistent with our burst timescale being $\sim$ the free-fall time at the edge of the disc), we predict a remaining cold gas fraction that is consistent with observations.
\end{enumerate}

Based on our results and discussion, we conclude there is clearly a significant stellar mass burst during galaxy mergers. Crucially important when comparing results is how the morphological selection of post-coalescence galaxies is performed. Samples relying on faint features could easily include minor mergers, which likely have a much smaller burst fraction than for major mergers.

Additionally, cold gas measurements, particularly of $\mathrm{H_2}$ for post-coalescence mergers, are needed to quantify the mass of cold gas remaining after a galaxy merger. As seen in portions of our work, derived burst fractions and timescales for post-starbursts are not always representative of all post-coalescence mergers. To measure $\mathrm{H_2}$ gas masses, a CO survey of at least one to two dozen post-coalescence galaxies, e.g. using a subset of the post-coalescence merger sample of \citet{Bickley2022}, is feasible with the Atacama Large Millimeter/submillimeter Array (ALMA).
%, but a similarly-sized survey may be a challenge for $\HI$ measurements done with current radio dish telescopes. The Square Kilometre Array will be able to quickly carry out these $\HI$ measurements as part of its all-sky survey, when it comes online in 2028/2029. Additionally, detailed mapping of the extended diffuse $\HI$ gas (beyond the galaxy disc), like that performed for a merging galaxy pair by \citet{Wang2023}, of a small sample of post-coalescence mergers with e.g. the Five-hundred meter Aperture Spherical Telescope (FAST) could greatly improve our understanding of the cooling of portions of the hot halo gas during galaxy mergers.

\section*{Acknowledgements}

We thank S.~Ellison for assistance with the SDSS post-coalescence merger catalogue. We also thank Liza Sazonova for informative discussions on post-starburst galaxies, Kristy Webb for helpful insights into the impact of particular choices of SED fitting routines, and Prathamesh Tamhane for discussions regarding hot halo physics.

The University of Waterloo acknowledges that much of our work takes place on the traditional territory of the Neutral, Anishinaabeg, and Haudenosaunee peoples. Our main campus is situated on the Haldimand Tract, the land granted to the Six Nations that includes six miles on each side of the Grand River.

%%%%%%%%%%%%%%%%%%%%%%%%%%%%%%%%%%%%%%%%%%%%%%%%%%
\section*{Data Availability}

The visually-confirmed post-coalescence merger sample of \citet{Bickley2022} is available for download from the MNRAS web version of the \citet{Bickley2022} publication. The SDSS data used in this work is publicly available as follows: Data Release 14 used in this work is available at \href{https://www.sdss4.org/dr14/data_access}{https://www.sdss4.org/dr14/data\_access}, and % catalogues with stellar masses are available from VizieR (\href{http://vizier.u-strasbg.fr}{vizier.u-strasbg.fr}), catalogue entry \texttt{J/MNRAS/379/867};
SFRs available from \href{https://wwwmpa.mpa-garching.mpg.de/SDSS/}{wwwmpa.mpa-garching.mpg.de/SDSS/}, %the two group catalogues are separately available through VizieR (catalogue entry \texttt{J/MNRAS/379/867}) and \href{http://gax.sjtu.edu.cn/ data/Group.html}{gax.sjtu.edu.cn/data/Group.html}. 
 mass- and luminosity-weighted ages and their accompanying stellar mass estimates are available from \href{https://www.sdss.org/dr16/spectro/galaxy\_firefly}{sdss.org/dr16/spectro/galaxy\_firefly}, and the Sérsic fits of \citet{Simard2011} come from table J/ApJS/196/11/table3 on \href{http://vizier.cfa.harvard.edu/}{VizieR}.



\bibliographystyle{mnras}
\bibliography{bibliography} % if your bibtex file is called example.bib

%%%%%%%%%%%%%%%%%%%%%%%%%%%%%%%%%%%%%%%%%%%%%%%%%%

%%%%%%%%%%%%%%%%% APPENDICES %%%%%%%%%%%%%%%%%%%%%

%\appendix

%%%%%%%%%%%%%%%%%%%%%%%%%%%%%%%%%%%%%%%%%%%%%%%%%%

% Don't change these lines
\bsp	% typesetting comment
\label{lastpage}
\end{document}